\documentclass[prb,twocolumn,groupedaddress,shownopacs,amssymb]{revtex4-2}
\usepackage{graphicx,psfrag,amsmath,amssymb}
\usepackage[usenames, dvipsnames]{color}
\usepackage{braket}
\usepackage{amsmath}
\usepackage{amssymb}
\usepackage{gensymb}
\usepackage{makeidx}
\usepackage{mathtools}
 
\usepackage[normalem]{ulem}
\usepackage{float}
\usepackage[T1]{fontenc} 
\usepackage{balance}
\usepackage{flushend}
\usepackage{lineno}
\usepackage{epsfig}
\usepackage{subfigure}
\usepackage{mathtools}
\usepackage{setspace}
\usepackage{bm}
\usepackage{ulem}
\usepackage{times}
\usepackage{enumitem}
\usepackage{xcolor}
\usepackage{multirow}
\usepackage{soul}
\usepackage{dcolumn}
\usepackage{balance}
\usepackage{epstopdf}
\usepackage{braket}
\usepackage{color}
\usepackage{diagbox}
\usepackage[utf8]{inputenc}
\usepackage{hyperref}
\usepackage{booktabs}
\usepackage{adjustbox}
\usepackage{makecell}
\hypersetup{
    colorlinks=true,
    linkcolor=Blue,
    citecolor=Blue,
    filecolor=Blue,
    urlcolor=Blue,
    }
\begin{document}
\title{Magnetic quantum phases of spin-orbit-coupled anisotropic dipolar bosons in square lattices}
\affiliation{Department of Physics, Central University of Rajasthan, Ajmer - 305817, India}
\author{Nitin Kaloya}
\affiliation{Department of Physics, Central University of Rajasthan, Ajmer - 305817, India}
\author{Kuldeep Suthar}
\affiliation{Department of Physics, Central University of Rajasthan, Ajmer - 305817, India}

\date{\today}
\begin{abstract}
 We examine the two-dimensional spin-orbit-coupled bosons in the presence of an anisotropic dipolar interaction in square lattices. The spin-orbit coupling leads to finite-momentum superfluid and
supersolid states, while the nearest-neighbour interaction induces crystalline characteristics in the quantum phases of soft-core bosons. We employ site-decoupled Gutzwiller ansatz and mean-field
decoupling theory to obtain the phase diagrams and investigate the effects of the tilt of magnetic dipoles with respect to the polarization axis. Our study reveals the intriguing quantum phase transition
of checkerboard finite-momentum phase-twisted and phase-stripe states into their stripe counterparts at a magic tilt angle, at which the off-site interaction along one of the directions becomes
zero. At smaller tilt angles, the checkerboard charge-density-wave phase intervened by two compressible finite-momentum phases, and at strong spin-orbit coupling strengths, the phase-twisted
supersolid and superfluid phases emerge. At larger tilt angles, a transition between the striped order of phase-twisted states and phase-stripe states occurs. The inclusion of off-site inter-component correlation leads to density-correlated phases, lattice-induced supersolid, and ferromagnetic quantum phases. Our study highlights novel finite-momentum crystal phases of spin-orbit-coupled dipolar bosons and provides a parameter space to observe them in quantum gas experiments. 
\end{abstract}
\maketitle

\section{Introduction}\label{intro}
Spin-orbit coupling (SOC), a coupling between the intrinsic angular momentum of a particle and its orbital motion, plays a significant role in determining the exotic phenomena of quantum condensed matter systems~\cite{zutic_04,lin_11,galitski_13,zhai_15}. The energy of spin-orbit coupling depends on the potential gradient and is strongly affected by the metallic surfaces. The role of coupling on the surface of the insulating states has gained a surge of interest due to possible implications for the topological insulator~\cite{hasan_10,liang_11,manchon_15}. In solids, the strength of spin-orbit coupling is an intrinsic property, and the interaction-driven process cannot be studied due to the lack of precise control over the electron density. The ultracold atoms in optical lattices provide an ideal platform to control the band structure, engineer synthetic spin-orbit coupling, and thus dress spin-dependent momentum of atoms through Raman
lasers~\cite{wu_16,lianghui_16,sun_18,hsun_19,curiel_21,wang_21,zejian_22,qian_24}. Moreover, the strength of SOC in ultracold experiments can be tuned to the limit, which is not possible in solids~\cite{galitski_13}. However, a strong correlation is also realized by considering the off-site density-density correlation in dipolar quantum gases. These systems are described by the extended Bose
Hubbard model (BHM), which is a minimal model and includes nearest-neighbour (NN) interaction of long-range anisotropic dipole-dipole interactions~\cite{lewenstein_07,boninsegni_12,chomaz_22}. The combined effects of spin-exchange complex tunneling and anisotropic dipolar interaction lead to rich spin textures, chiral supersolid (SS), topological defects, and quantum quasicrystal states~\cite{deng_12,gopalkrishnan_13,wilson_13,wei_18}. 

Even though there are plenty of studies on the effects of long-range dipolar interaction and synthetic spin-orbit coupling in ultracold quantum gases in optical lattices, the combined interplay of these two has remained relatively less explored. The nearest-neighbour off-site interaction introduces a density-correlated supersolid phase, which is stabilized for soft-core bosons due to larger Fock space and number fluctuations, in contrast to hard-core constraints~\cite{otterlo_95,batrouni_00,sengupta_05,sansone_10,ohgoe_12}. Moreover, the canting (anisotropic nature) of the dipoles out of the lattice plane by an external electric field results in a quantum phase transition from checkerboard to stripe orders through emulsion phases~\cite{zhang_15,bandyopadhyay_19,wu_20,zhang_21,zhang_22,nguyen_22}. Recent experiments have realized the supersolid phase~\cite{leonard_17,li_17,guo_19,tanzi_19,natale_19,norcia_21,bland_22,kirkby_24} and checkerboard spatial order~\cite{lagoin_22} and the role of tilt of dipoles in the superfluid-insulator transition~\cite{baier_16}. In addition, supersolidity due to anomalous weak dipolar interaction~\cite{wilson_16}, cavity-mediated long-range interactions~\cite{guan_19}, and phase separation phenomena of the supersolid phase~\cite{bai_20} have also been studied. Besides, the synthetic SOC causes magnetic phase transitions from stripe to magnetized plane-wave as the first transition and plane-wave to non-magnetized zero momentum as the second phase transition~\cite{ji_14}. Furthermore, the two-dimensional system exhibits topological phase transitions due to inversion and $C_{4}$ rotational symmetries~\cite{sun_18}. The synthetic SOC has also been created in dipolar gases~\cite{burdick_16,norcia_21a}. On the theoretical front, the ferromagnetic and antiferromagnetic phases~\cite{gong_15}, spin-spiral orders~\cite{cai_12,zhang_19}, and vortex and skyrmion crystal phases~\cite{cole_12,sousa_25} of spin-orbit-coupled lattice bosons have been examined. The optical lattice potential leads to an improvement in the contrast and lifetime of the stripe superfluid (SF) state~\cite{bersano_19}. In addition to the rich spin structures, the SOC breaks the charge order of the insulating states and leads to finite-momentum phase-twisted (PT) and phase-stripe (PS) superfluids~\cite{grass_11,mandal_12,hickey_14,dutta_19}. More recently, the parameter space of these novel soft-core boson superfluids and the effects of NN interactions have been investigated~\cite{suthar_21,pu_24}. However, the role of anisotropic dipolar interaction in the emergence of exotic orders of finite-momentum superfluids and supersolids has not been explored.

In this work, we investigate the role of tilt of soft-core dipolar bosons in the presence of synthetic spin-orbit coupling. To this end, we employ the site-decoupled Gutzwiller approach and mean-field decoupling theory to obtain the phase diagrams of anisotropic dipolar bosons in two-dimensional optical lattices. Here, we identify the ordering of the correlated states induced in the amplitude and phase of the order parameter due to NN interaction and spin-orbit coupling, respectively. We show the transition from checkerboard phase-twisted and phase-stripe to their respective striped order as a function of tilt angle through emulsion states. The interspin off-site correlation admits correlated lattice-induced supersolidity and phase-separated ferromagnetic states. 

The paper is organized as follows: In Sec.~\ref{ham} we introduce the model Hamiltonian of spin-orbit-coupled dipolar bosons in square lattices and provide a brief description of the mean-field Gutzwiller approach. In Sec.~\ref{class_phases}, we discuss the characterization of quantum phases and order parameters. Section~\ref{mfdt} discusses the mean-field decoupling theory to obtain the phase boundaries of compressible-incompressible phase transitions of anisotropic dipolar bosons. The phase diagrams of anisotropic dipolar bosons with different SOC strengths are discussed in Sec.~\ref{res}. Finally, we conclude in Sec.~\ref{conc}.

\section{Model and Methods}
\label{ham}
We consider a pseudospinor two-component ultracold dipolar bosons trapped in a two-dimensional square optical lattices. The Raman beams couple atomic momentum with spin and generate synthetic Rashba spin-orbit coupling, experienced by dipolar bosons. The system is described by the two-component extended Bose-Hubbard model for anisotropic dipoles in the presence of spin-orbit coupling. The model Hamiltonian of the system is  
\begin{eqnarray}
 \hat{H} &=& - \sum_{p,q,\sigma} \biggr[\left(J_x \hat{b}^{\dagger\sigma}_{p+1,q} \hat{b}^{\sigma}_{p,q} 
           + J_y \hat{b}^{\dagger\sigma}_{p,q+1} \hat{b}^{\sigma}_{p,q} + \text{H.c.} \right) \nonumber \\ 
          &-& \frac{U_{\sigma\sigma}}{2} \hat{n}^{\sigma}_{p,q}\left(\hat{n}^{\sigma}_{p,q}-1\right) 
           + {\mu}^{\sigma}_{p,q} \hat{n}^{\sigma}_{p,q} \biggr] \nonumber \\
          &+& \sum_{p,q} U_{\uparrow\downarrow} \hat{n}^{\uparrow}_{p,q} \hat{n}^{\downarrow}_{p,q} 
           + \hat{H}_{\rm soc} + \hat{H}_{V},
\label{model_ham}
\end{eqnarray}
where the Hamiltonian term corresponding to spin-orbit coupling is
\begin{eqnarray}
\hat{H}_{\rm{soc}} &=&-\gamma_x \sum_{p,q} 
 \Big[ \hat{b}^{\uparrow\dagger}_{p,q} \hat{b}^{\downarrow}_{p+1,q} 
     - \hat{b}^{\downarrow\dagger}_{p,q} \hat{b}^{\uparrow}_{p+1,q} \Big]  \nonumber \\
&-& i \gamma_y \sum_{p,q} 
 \Big[ \hat{b}^{\uparrow\dagger}_{p,q+1} \hat{b}^{\downarrow}_{p,q} 
     + \hat{b}^{\downarrow\dagger}_{p,q+1} \hat{b}^{\uparrow}_{p,q} \Big] + \text{H.c.},
\end{eqnarray} 
and the last term is the anisotropic nearest-neighbour dipolar interaction
\begin{eqnarray}
\hat{H}_{V} &=& \frac{1}{2} \sum_{p,q,\sigma} \Big[ V^{x}_{\sigma\sigma} \hat{n}^{\sigma}_{p,q} \big(\hat{n}^{\sigma}_{p+1,q} 
                                               + \hat{n}^{\sigma}_{p-1,q}\big) \nonumber \\
&+& V^{y}_{\sigma\sigma}\,\hat{n}^{\sigma}_{p,q}\big(\hat{n}^{\sigma}_{p,q+1} + \hat{n}^{\sigma}_{p,q-1}\big) \Big] \nonumber \\
&+& \sum_{p,q} \Big[V^{x}_{\uparrow\downarrow} \hat{n}^{\sigma}_{p,q}\big(\hat{n}^{\sigma'}_{p+1,q} + \hat{n}^{\sigma'}_{p-1,q}\big) \nonumber \\
&+& V^{y}_{\uparrow\downarrow}\, \hat{n}^{\sigma}_{p,q}\big(\hat{n}^{\sigma'}_{p,q+1} + \hat{n}^{\sigma'}_{p,q-1}\big) \Big].
\end{eqnarray}
Here, $\sigma = \{\uparrow, \downarrow\}$ labels the two pseudospin components and $(p,q)$ denotes the lattice site indices with $p$ and $q$ are 
site indices along $x$ and $y$ directions, respectively. The operator $\hat{b}_{p,q}^{\dagger\sigma}$ ($\hat{b}_{p,q}^{\sigma}$) creates (annihilates) a boson of $\sigma$-component at site $(p,q)$ and $\hat{n}_{p,q}^{\sigma} = \hat{b}_{p,q}^{\dagger\sigma}\hat{b}_{p,q}^{\sigma}$ is the number operator for $\sigma$-component at site $(p,q)$. The parameter $\mu_{p,q}^{\sigma}$ is the local chemical potential for the spin-component $\sigma$ and it is considered to be same for both of the components. The parameters $J_x$ ($J_y$) and $\gamma_x$ ($\gamma_y$) are the spin-independent hopping amplitudes and spin-orbit coupling (spin-dependent complex hopping) strength along the $x$ ($y$) direction, respectively. We assume identical hopping strengths in both directions, $J_x = J_y = J$ and $\gamma_x = \gamma_y = \gamma$. The on-site intraspin interaction is characterized by $U_{\sigma\sigma}$, whereas $U_{\uparrow\downarrow}$ represents the interspin interaction. These interaction strengths depend on the optical lattice depth, the recoil energy, and the atomic $s$-wave scattering lengths. Furthermore, $V^{x}_{\sigma\sigma}$ ($V^{y}_{\sigma\sigma}$) and $V^{x}_{\uparrow,\downarrow}$ ($V^{y}_{\uparrow,\downarrow}$) are intra- and interspin NN off-site interactions. The intraspin on-site and NN interactions are also assumed to be identical, $U_{\uparrow\uparrow} = U_{\downarrow\downarrow} = U$ and $V_{\uparrow\uparrow} = V_{\downarrow\downarrow} = V$. The intercomponent NN interaction is parameterized as $V_{\uparrow\downarrow} = \zeta V$. 
\begin{figure}[h]
\includegraphics[width=\linewidth]{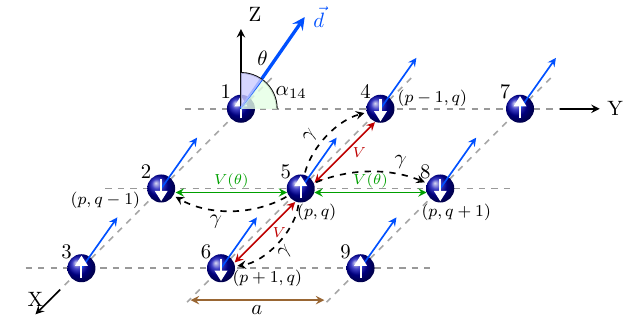}
\caption{Schematic representation of spin-flip complex hopping ($\gamma$) of spin-orbit coupled dipolar bosons in two-dimensional square optical lattices. The dipoles are assumed to lie within the $yz$-plane and the angle $\theta$ subtended by the polarization axis with the $z$-axis is the tilt angle, that defines the direction of polarization. The tilt angle is shown by the blue-shaded region. The angle between the polarization axis and the relative position vector $(r_4 - r_1)$, $\alpha_{14}$, is shown by the green shaded region. The dipolar interaction strengths between bosons at lattice sites $(p,q)$ and $(p \pm 1,q)$ are denoted by $V_{\sigma \sigma}$ and $V_{\uparrow \downarrow}$ for intra- and interspin interactions, whereas these interactions along y-direction at $(p,q)$ and $(p,q \pm 1)$ depend on the tilt angle with $V_{\sigma \sigma}(\theta) = V_{\sigma \sigma}(1 - 3 \sin^2{\theta})$ and $V_{\uparrow \downarrow}(\theta) = V_{\uparrow \downarrow}(1 - 3 \sin^2{\theta})$. Here, $a$ is the lattice constant.}
\label{spin_up_down}
\end{figure}

The nearest-neighbour off-site interaction between bosons stems from the dipole-dipole interaction, and expressed as  
$\dfrac{C_{\mathrm{dd}}}{2} \sum_{i,j} \hat{n}_{i} \hat{n}_{j} \dfrac{(1 - 3 \cos^2 \alpha_{ij})}{|\mathbf{r}_{i} - \mathbf{r}_{j}|^3}$, where $\alpha_{ij}$ is the angle between the dipole polarization axis and the vector connecting sites $i$ and $j$, and $C_{\mathrm{dd}}$ denotes the strength of dipolar interaction ~\cite{bai_20, bandyopadhyay_19}. The polarization angle $\alpha_{ij}$ can be experimentally tuned by rotating the dipoles using a time-dependent external magnetic field ~\cite{giovanazzi_02,tang_18}. Although the dipolar interaction is inherently long-ranged, in the present work we restrict it to the nearest-neighbour sites, effectively considering only NN anisotropic interactions. The nearest-neighbour interaction can act either within a single spin-component or between different spin components, corresponding to intraspin ($V_{\sigma\sigma}$) and interspin ($V_{\uparrow\downarrow}$) interactions, respectively.

The orientations of dipoles that are polarized in the $yz$ plane is shown in Fig.~\ref{spin_up_down}. The spin-exchange in the hopping occurs because of synthetic spin-orbit coupling. Note that the spin-independent single-particle hopping is not shown in the figure. The angle $\alpha_{ij}$ varies with $\theta$ and can be tuned by changing the direction of the applied magnetic field. The intra- and interspin interactions along the $x$ direction is always constant, repulsive, and independent of $\theta$, whereas along the $y$ direction it depends on $\theta$ as $V_{\sigma \sigma}(\theta)=V_{\sigma \sigma}(1-3\sin^2\theta)$ and $V_{\uparrow \downarrow}(\theta) = V_{\uparrow \downarrow}\left(1 - 3\sin^2\theta\right)$. NN interactions change from $V$ at $\theta=0^\circ$ to $-2V$ at $\theta=90^\circ$, and vanish at the magic angle $\theta_M=\sin^{-1}(1/\sqrt{3})\approx35.3^\circ$. Thus, the interaction along the $y$ direction is repulsive for $\theta<\theta_M$ , while for $\theta>\theta_M$, it becomes attractive.

The single-particle kinetic Hamiltonian in momentum space can be written as
\begin{equation*}
  \hat{H}_{\mathrm{kin}}
  = \sum_{\mathbf{k}}
  \begin{pmatrix}
  \hat{b}^{\uparrow \dagger}_{\mathbf{k}} &
  \hat{b}^{\downarrow \dagger}_{\mathbf{k}}
  \end{pmatrix}
  \mathcal{H}_{\mathbf{k}}
  \begin{pmatrix}
  \hat{b}^{\uparrow}_{\mathbf{k}} \\
  \hat{b}^{\downarrow}_{\mathbf{k}}
  \end{pmatrix},
\end{equation*}
where
\begin{equation*}
\mathcal{H}_{\mathbf{k}}
= -2J(\cos k_x + \cos k_y)\,\hat{I}
+ 2\gamma\left(\sin k_y\,\hat{\sigma}_x - \sin k_x\,\hat{\sigma}_y\right).
\end{equation*}
The corresponding single-particle energy dispersion exhibits two branches
\begin{equation*}
E_{\mathbf{k}}^{\pm}
= -2J(\cos k_x + \cos k_y)
\pm 2\gamma \sqrt{\sin^2 k_x + \sin^2 k_y}.
\end{equation*}
These bands exhibit four degenerate minima located at ${Q_1} = (k_0,k_0)$, ${Q_2} = (-k_0,k_0)$, ${Q_3} = (-k_0,-k_0)$, and ${Q_4} = (k_0,-k_0)$
where $k_0 = \arctan\!\left(\frac{\gamma}{\sqrt{2}\, J}\right)$.

Various quantum phase transitions of a two-dimensional spin–orbit coupled BHM with nearest-neighbour interactions [Eq.~(\ref{model_ham})] are studied using site-decoupled Gutzwiller mean-field theory under periodic boundary conditions. In this approach, the operators are expressed as an average around fluctuating operator $\hat{b}_{i} \rightarrow \langle\hat{b}_{i}\rangle + \delta\hat{b}_{i}$ $(i=\{p,q\})$, where $\langle\hat{b}_{i}\rangle \equiv \phi_{i}$ is the mean-field superfluid order parameter and $\delta\hat{b}_{i}$ is the fluctuation operator. With this approximation, the Hamiltonian decouples the sites, and all the off-site contributions are incorporated through the mean field. The many-body Gutzwiller wave function is~\cite{gutzwiller_63,rokshar_91,sheshadri_93,bai_18,bandyopadhyay_19,suthar_20,bai_20,suthar_21,suthar_22} 
\begin{equation}
|\Psi\rangle = \prod_{i} \sum_{n_{\uparrow}, n_{\downarrow}}^{n_{\text{max}}}
c^{i}_{n_{\uparrow},n_{\downarrow}} \, |n_{\uparrow}, n_{\downarrow}\rangle_{i},
\label{gutz_mb}
\end{equation}
where $|n_{\uparrow}, n_{\downarrow}\rangle_{i}$ denotes the local Fock state at $i$th lattice site, and $n_{\sigma} = 0, 1, \dots,n_{\text{max}}$ is the occupation number of spin-$\sigma$ bosons, with $n_{\text{max}}$ being the truncation limit. The complex variational coefficients $c^{i}_{n_{\uparrow},n_{\downarrow}}$ satisfy the normalization condition $\sum_{n_{\uparrow}, n_{\downarrow}}^{n_{\text{max}}} 
|c^{i}_{n_{\uparrow}, n_{\downarrow}}|^2 = 1$. The many-body wave-function [Eq.~(\ref{gutz_mb})] is the product of single-site states such that 
the operators of the Hamiltonian Eq.~(\ref{model_ham}) which are over NN sites can be written in terms of on-site operators. For single-particle hopping, complex hopping, and NN interactions involving sites $i$ and $j$, these are given as 
\begin{subequations}
\begin{eqnarray}
\hat{b}^{\dagger\sigma}_{i} \hat{b}^{\sigma'}_{j} &=& \langle \hat{b}^{\dagger\sigma}_{i} \rangle \hat{b}^{\sigma'}_{j} + \hat{b}^{\dagger\sigma}_{i} \langle \hat{b}^{\sigma'}_{j} \rangle - \langle \hat{b}^{\dagger\sigma}_{i} \rangle \langle \hat{b}^{\sigma'}_{j} \rangle, \\
\hat{n}^{\sigma}_{i} \hat{n}^{\sigma}_{j} &=& \langle \hat{n}^{\sigma}_{i} \rangle \hat{n}^{\sigma}_{j} + \hat{n}^{\sigma}_{i} \langle \hat{n}^{\sigma}_{j} \rangle - \langle \hat{n}^{\sigma}_{i} \rangle \langle \hat{n}^{\sigma}_{j} \rangle.
\end{eqnarray}
\end{subequations}
The superfluid order parameters characterizing single-atom transport and coherence for the two spin components are
\begin{subequations}
\begin{eqnarray}
\phi^{\uparrow}_{i} &=& \langle \Psi | \hat{b}^{\uparrow}_{i} | \Psi \rangle = \sum_{n_{\uparrow}, n_{\downarrow}}^{n_{\text{max}}} 
\sqrt{n^{\uparrow}_{i}}\ c^{*i}_{n_{\uparrow}-1, n_{\downarrow}}\ c^{i}_{n_{\uparrow}, n_{\downarrow}}, \\
\phi^{\downarrow}_{i} &=& \langle \Psi | \hat{b}^{\downarrow}_{i} | \Psi \rangle = \sum_{n_{\uparrow}, n_{\downarrow}}^{n_{\text{max}}} 
\sqrt{n^{\downarrow}_{i}} \, c^{*i}_{n_{\uparrow}, n_{\downarrow}-1} c^{i}_{n_{\uparrow}, n_{\downarrow}}
\end{eqnarray}
\end{subequations}
and the corresponding average on-site densities are given by
\begin{subequations}
\begin{eqnarray}
\rho^{\uparrow}_{i} &=& \langle \Psi | \hat{b}^{\uparrow\dagger}_{i} \hat{b}^{\uparrow}_{i} | \Psi \rangle 
= \sum_{n_{\uparrow}, n_{\downarrow}}^{n_{\text{max}}} 
n_{i}^{\uparrow} \, \left| c^{i}_{n_{\uparrow}, n_{\downarrow}} \right|^2, \\
\rho^{\downarrow}_{i} &=&
\langle \Psi | \hat{b}^{\downarrow\dagger}_{i} \hat{b}^{\downarrow}_{i} | \Psi \rangle 
= \sum_{n_{\uparrow}, n_{\downarrow}}^{n_{\text{max}}} 
n_{i}^{\downarrow} \, \left| c^{i}_{n_{\uparrow}, n_{\downarrow}} \right|^2.
\end{eqnarray}
\end{subequations}
Due to the presence of spin–orbit coupling, the SF order parameter at the $j$th site becomes complex and can be expressed as $\phi^{\sigma}_{j} = \left| \phi^{\sigma}_{j} \right| e^{i \theta^{\sigma}_{j}}$, where $\left| \phi^{\sigma}_{j} \right|$ and $\theta^{\sigma}_{j}$ represent its magnitude and phase, respectively. Since the spin-up and spin-down components are related by exchange symmetry, the magnitudes of their order parameters are equal.

\section{Characterization of Quantum Phases}
\label{class_phases}
Table~\ref{table_intra} summarizes the ground state quantum phases of two-dimensional spin-orbit coupled (extended) Bose-Hubbard model with nearest-neighbour interactions for canted dipolar bosons. In the conventional two-component BHM, the ground states are either a Mott insulator (MI) or a superfluid phase. The MI is an incompressible phase and characterized by a vanishing superfluid order parameter, $|\phi_i^{\sigma}| = 0$, with an integer and spatially uniform density $n_i^{\sigma} = n_j^{\sigma} \in \mathbb{N}$ (with $i$ and $j$ being sublattice indices). In contrast, the superfluid is a compressible phase with a finite SF order parameter, $|\phi_i^{\sigma}| \neq 0$, and a uniform but non-integer density $n_i^{\sigma} = n_j^{\sigma} \in \mathbb{R}$. For both MI and SF phases, the density remains uniform across the lattice, resulting in vanishing density imbalance, $\Delta n^{\sigma} = n_i^{\sigma} - n_j^{\sigma} = 0$.
\begin{table}[h!]
\begin{ruledtabular}
\begin{tabular}{l c c c c }
\makecell{\text{Quantum} \\ \text{phase}} & $\Delta n^{\sigma} (n_i^\sigma)$ & $\Delta\phi_i^\sigma (\phi_i^\sigma)$ & \makecell{$S(\pi,0)$ $\small/$ \\ $S(0,\pi)\phantom{/}$} & $S(\pi,\pi)$ \\
\colrule
CBDW & $\neq 0$ (Integer) & $ = 0$ (0) & $ = 0$ & $\neq 0$ \\
SDW  & $ = 0$ (Integer) & $ = 0$ (0) & $\neq 0$    & $ = 0$   \\
EDW  & $\neq 0$ (Integer) & $ = 0$ (0) & $\neq 0$    & $\neq 0$  \\
cDW\footnote{The cDW phase can be distinguished from the EDW phase based on the spatial distribution of the individual component occupancies.} 
     & $\neq 0$ (Integer) & $ = 0$ (0) & $\neq 0$ & $\neq 0$  \\
cMI  & $= 0$ (Integer) & $ = 0$ (0) & $= 0$ & $= 0$ \\
zFM\footnote{The zFM phase can be distinguished from the SF phase on the basis of magnetization $\mathcal{M}$, which is finite in the zFM phase and vanishes in the SF phase.}
     & $= 0$ (Real) & $ = 0$ ($\neq 0$) & $= 0$ & $= 0$  \\
SF   & $ = 0$ (Real) & $ = 0$ ($\neq 0$) & $ = 0$ & $ = 0$  \\
ESS  & $\neq 0$ (Real) & $\neq 0$ ($\neq 0$) & $\neq 0$ & $\neq 0$  \\
SSS  & $\neq 0$ (Real) & $\neq 0$ ($\neq 0$) & $\neq 0$ & $ = 0$  \\
DCSS & $\neq 0$ (Real) & $\neq 0$ ($\neq 0$) & $ = 0$ & $\neq 0$  \\
LSS\footnote{The LSS phase can be distinguished from the DCSS phase based on the spatial distribution of the individual component occupancies.}
     & $\neq 0$ (Real) & $\neq 0$ ($\neq 0$) & $ = 0$ & $\neq 0$ \\
\end{tabular}
\end{ruledtabular}
	\caption{Classification of phases: Sublattice values of the occupations $n_i^\sigma$ and SF order parameters $\phi_i^\sigma$ with corresponding sublattice distribution difference. Here, $S(\mathbf{k})$ is the structure factor at reciprocal lattice vector $\mathbf{k}$.}
\label{table_intra}
\end{table}
When NN interactions are included, two additional quantum phases appear: the density-wave (DW) phase and the supersolid phase. These phases exhibit spatially nonuniform densities, i.e., $\Delta n^{\sigma} \neq 0$. The DW is an incompressible phase with $|\phi_i^{\sigma}| = 0$, and the site occupations on the two sublattices take different integer values, $n_i^{\sigma}, n_j^{\sigma} \in \mathbb{N}$, giving rise to $\Delta n^{\sigma} \in \mathbb{N}$. In general, the density of each component forms a checkerboard pattern, while the total density remains uniform, satisfying $\Delta n^{\uparrow} = -\Delta n^{\downarrow}$. The SS is a compressible phase and features finite, spatially modulated SF order parameters, $|\phi_i^{\sigma}| \neq 0$ and $|\phi_i^{\sigma}| \neq |\phi_j^{\sigma}|$. In addition, a double-checkerboard supersolid (DCSS) phase emerges when only intraspin NN interactions are present with $\zeta = 0$. The DCSS phase has real densities with $n_i^{\sigma} \neq n_j^{\sigma}$ and displays a checkerboard pattern for each component, while the total density remains uniform. When both intra- and interspin nearest-neighbour interactions are present, the system realizes a correlated lattice supersolid (LSS) phase. Depending on the strength of SOC, the system exhibits PT-LSS, PS-LSS, and ZM-LSS phases. The occupancies of the individual components, their combined occupancy, and the corresponding phase $\theta^{\uparrow}$ are illustrated in Fig.~\ref{phase_lss}.

\subsection{Spin-dependent momentum}
The spin-orbit coupling leads to variations in phases of compressible states and leads to novel finite-momentum superfluids namely phase-twisted and phase-stripe superfluid states, characterized by phase structures of order parameters. These states are distinguished by the spin-dependent momentum \cite{dutta_19,suthar_21} 
\begin{equation}
  \langle \rho_{\uparrow\downarrow}(\mathbf{k}) \rangle
  = \frac{1}{L^2} \sum_{i,j}
  \langle \hat{b}^{\uparrow\dagger}_{i}\, \hat{b}^{\downarrow}_{j} \rangle
  \, e^{i\mathbf{k}\cdot(\mathbf{r}_i - \mathbf{r}_j)} ,
\end{equation}

where $L$ is system size and $\mathbf{r}_i - \mathbf{r}_j$ is the distance between the $i$th and $j$th sites. In the presence of the interactions, the SF states can be divided into three types. In the homogeneous superfluid phase, the order parameter maintains a constant magnitude and phase throughout the lattice. In this regime, Bose condensation occurs at zero momentum. In the PT superfluid, the magnitude of the local order parameter $\langle \hat{b}_{i\sigma} \rangle$ remains spatially uniform, while its phase exhibits a  systematic 
variation along the diagonal directions of the lattice. Consequently, the $\langle \rho_{\uparrow\downarrow}(\mathbf{k}) \rangle$ develops a peak at one of the characteristic diagonal momenta $Q$'s. The momentum distribution then becomes concentrated along the diagonal of the first Brillouin zone, specifically at $\langle \rho_{\uparrow\downarrow}(-k_0,-k_0) \rangle$ or $\langle \rho_{\uparrow\downarrow}(k_0,k_0) \rangle$.
\begin{figure}[h]
    \includegraphics[width=\linewidth]{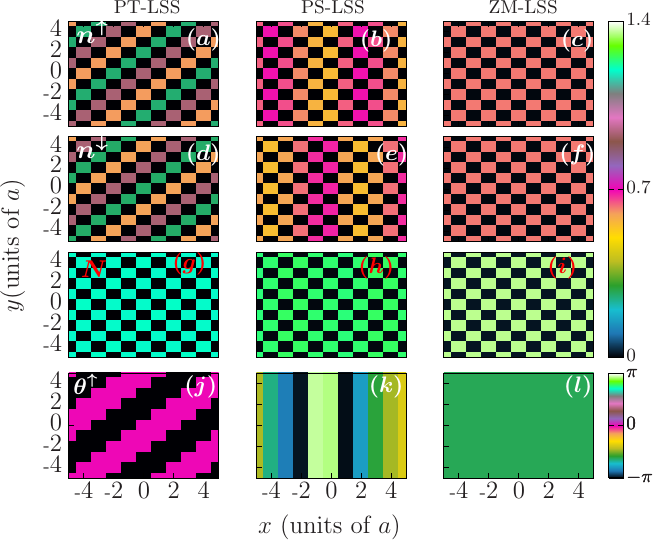}
    \caption{Spatial distributions of spin-resolved densities and phases in lattice supersolid states. Panels $(a)\!\!-\!\!(c)$ show the spin-up density $n^{\uparrow}$, $(d)\!\!-\!\!(f)$ the spin-down density $n^{\downarrow}$, and $(g)\!\!-\!\!(i)$ the total density $N = n^{\uparrow} + n^{\downarrow}$. Panels $(j)\!-\!(l)$ display the phase $\theta^{\uparrow}$ of spin-up component. The three columns correspond to the phase-twisted lattice SS, phase-striped lattice SS, and zero-momentum lattice SS phases, respectively. Here $a$ is the lattice constant.}
    \label{phase_lss}
\end{figure}
\begin{figure}[h]
    \includegraphics[width=\linewidth]{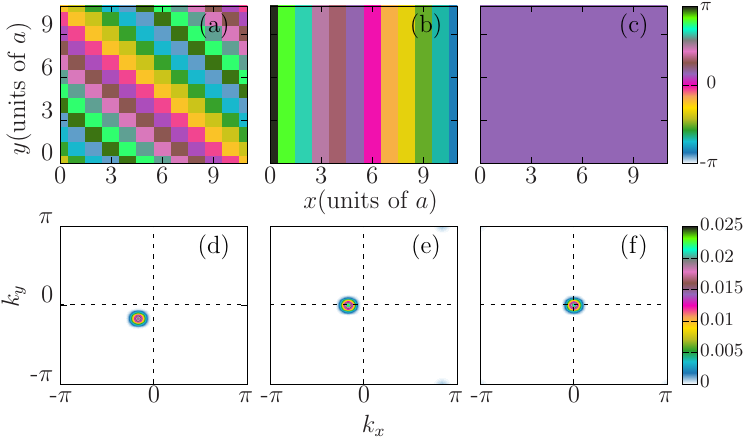}
    \caption{Real-space phase $\theta^{\sigma}$ variations across lattice and the associated spin-resolved momentum distributions for different supersolid states. Upper panel (a,b,c) depicts the phase configurations of the PT, PS, and ZM supersolid, while lower panel (d,e,f) displays the corresponding momentum-space distributions. Here, $a$ is the lattice constant.}
    \label{ss_ph}
\end{figure}
In contrast, in the PS-SF phase, the phase varies along either the horizontal or vertical lattice directions. As a result, the momentum distribution exhibits peaks at $\langle \rho_{\uparrow\downarrow}(\pm k_0,0) \rangle$ or $\langle \rho_{\uparrow\downarrow}(0,\pm k_0) \rangle$, corresponding to stripe-like modulations of the condensate. Similarly, within the supersolid region, the presence of SOC enriches the phase diagram and gives rise to the phase-twisted supersolid (PT-SS) and phase-striped supersolid (PS-SS) states. The characteristic properties of the magnetic supersolid phases are shown in Fig.~\ref{ss_ph}. Moreover, the LSS states also exhibits similar phase distributions [Fig.~\ref{phase_lss}(j,k,l)]. We introduce the quantity $\varrho = \langle \rho_{\uparrow\downarrow}(k_{0},0) \rangle + \langle \rho_{\uparrow\downarrow}(-k_{0},0) \rangle + \langle \rho_{\uparrow\downarrow}(0,k_{0}) \rangle + \langle \rho_{\uparrow\downarrow}(0,-k_{0}) \rangle,$ which acts as an order parameter for identifying the PT-PS superfluid phase transition. In the PT superfluid phase, this parameter is zero, while in the PS superfluid phase it takes a finite value. Moreover, for higher value of hopping amplitude, the zero-momentum SF (ZM-SF) with the spin-dependent momentum peak $\langle \rho_{\uparrow \downarrow}(k)$ at $\mathbf{k}=0$ emerges when $U_{\uparrow \downarrow}/U < 1$. Thus, the distinct finite-momentum superfluid states can be classified by examining the condensate phase distribution across the lattice and by analyzing their respective momentum-space measures. 


\subsection{Static structure factor}
With the long-range dipole-dipole NN interactions, the system can spontaneously breaks translational symmetry. This gives rise to phases with spatially modulated densities; most notably, density-wave and supersolid phases, reflecting the presence of diagonal long-range order. The SS phase displays both diagonal order and off-diagonal long-range order (ODLRO); hence it features a nonzero $\phi^{\sigma}_{i}$ along with a periodically varying density $n^{\sigma}_{i}$. To characterize the diagonal order of the phases, we evaluate the static structure factor~\cite{otterlo_94,batrouni_00}
\begin{equation}
    S(\mathbf{k}) = \frac{1}{L^2} \sum_{i,j} e^{i\mathbf{k}\cdot(\mathbf{r}_i - \mathbf{r}_j)} 
\langle \hat{n}^{\sigma}_i \hat{n}^{\sigma}_j \rangle.
\end{equation}
Depending on the tilt angle $\theta$ of dipoles, the density configuration $n^{\sigma}_{i}$ may adopt either a checkerboard or a striped pattern. Checkerboard ordering breaks translational symmetry in both lattice directions and is signaled by a finite value of $S(\pi,\pi)$. In striped phases, symmetry breaking occurs only along one direction, and the corresponding signature appears in $S(\pi,0)$ or $S(0, \pi)$. Thus, $S(\pi,\pi)$, $S(\pi,0)$ and $S(0, \pi)$ effectively distinguish checkerboard (CB) and stripe-ordered states, respectively. Near magic angle both orders compete and the system resides in emulsion incompressible and compressible states having both finite $S(\pi,\pi)$ and $S(\pi,0)$.

\section{Mean-field decoupling theory}
\label{mfdt}
We now turn to discuss the analytical mean-field decoupling theory to obtain the phase boundaries between incompressible and compressible phases. This approach has been previously employed to obtain the phase boundaries of BHM~\cite{oosten_01,iskin_09} and eBHM~\cite{bandyopadhyay_19,bai_20,pu_24}. In this perturbative analysis, as the incompressible phases have zero SF order parameter, the critical hopping strength of transitions can be marked by $\phi^{\sigma}_{i}\rightarrow 0^{+}$. For the canted spin-orbit coupled dipolar bosons, the unperturbed ground-state energy at $i$ and $j$ sublattices is
\begin{eqnarray}
    E^{(0)}_{n_i^\uparrow, n_i^\downarrow; \, n_j^\uparrow, n_j^\downarrow}
    &=&\sum_\sigma \biggr[
        \frac{U}{2} n_i^\sigma (n_i^\sigma - 1)
        + V^{x}_{\sigma\sigma} n_i^\sigma n_j^\sigma \nonumber \\
    && + V^{y}_{\sigma\sigma}(\theta)n_i^\sigma n_j^\sigma 
        - \mu n_i^\sigma
    \biggr] + U_{\uparrow\downarrow} n_i^\uparrow n_i^\downarrow \nonumber \\
    &&+ V^{x}_{\uparrow \downarrow}(n_{i}^{\uparrow} n_{j}^{\downarrow} + n_{i}^{\downarrow} n_{j}^{\uparrow}) \nonumber \\
    &&+ V^{y}_{\uparrow \downarrow}(\theta)(n_{i}^{\uparrow} n_{j}^{\downarrow} + n_{i}^{\downarrow} n_{j}^{\uparrow}),  
\label{ener_up}
\end{eqnarray}
 To investigate the phase boundary between the incompressible and compressible phases, we treat the single-particle hopping amplitude and the spin-orbit coupling terms as perturbations. 
We first discuss the quantum phase transitions below the magic tilt angle, where the phases possess checkerboard order. Within the mean-field approximation, the perturbed Hamiltonian $\hat{H}^{c}_{\rm per}$ is expressed as
\begin{equation}
\begin{aligned}
\hat{H}^{c}_{per} = &-J \sum_{\sigma} \Big[ \phi_i^{\sigma} \left( \hat{b}_j^{\sigma \dagger} + \hat{b}_j^{\sigma} \right) + \phi_j^{\sigma} \left( \hat{b}_i^{\sigma \dagger} + \hat{b}_i^{\sigma} \right) \Big] \\
&+ \gamma_x \Big[ \bar{\phi}_i^{x \downarrow} \left( \hat{b}_{i}^{\uparrow \dagger } -\hat{b}_{i}^{\uparrow} \right) + \bar{\phi}_i^{x\uparrow} \left( \hat{b}_{i}^{\downarrow \dagger} - \hat{b}_{i}^{\downarrow} \right) \Big] \\
&+ \gamma_x \Big[ \bar{\phi}_j^{x \downarrow} \left( \hat{b}_{j}^{ \uparrow \dagger} -\hat{b}_{j}^{\uparrow} \right) + \bar{\phi}_j^{x\uparrow} \left( \hat{b}_{j}^{\downarrow \dagger } - \hat{b}_{j}^{\downarrow} \right) \Big] \\
&+ i\gamma_y \Big[ \bar{\phi}_i^{y \downarrow} \left( \hat{b}_{i}^{\uparrow \dagger}-\hat{b}_{i}^{\uparrow} \right) + \bar{\phi}_i^{y \uparrow} \left( \hat{b}_{i}^{\downarrow \dagger}-\hat{b}_{i}^{\downarrow} \right) \Big] \\
&+ i\gamma_y \Big[ \bar{\phi}_j^{y \downarrow} \left( \hat{b}_{j}^{\uparrow \dagger}-\hat{b}_{j}^{\uparrow} \right) + \bar{\phi}_j^{y \uparrow} \left( \hat{b}_{j}^{ \downarrow \dagger}-\hat{b}_{j}^{\downarrow} \right) \Big].
\end{aligned}
\label{ham_per}
\end{equation}
Here, $\bar{\phi}_{i}^{\sigma} = \sum_{\langle j\rangle_{i}} \phi_{j}^{\sigma}$ and
$\bar{\phi}_{j}^{\sigma} = \sum_{\langle i\rangle_{j}} \phi_{i}^{\sigma}$ represent the sums of the order parameters over all nearest-neighbour sites $j$ of site $i$ and all nearest-neighbour sites $i$ of site $j$, respectively.

Using perturbative analysis, we obtain the dependence of the critical hopping $J_c$ on the SOC strength $\gamma$. The details are discussed in the appendix~\ref{sec-I}. In the presence of only intraspin nearest-neighbour interactions with $\zeta = 0$, the occupation numbers satisfy $n_i^{\uparrow} = n_j^{\downarrow}$ and $n_i^{\downarrow} = n_j^{\uparrow}$ for the checkerboard density wave (CBDW) phase. The minimization of second-order corrected ground-state energy results in the following equation, which determines the CBDW-SF (SS) phase boundary
\begin{eqnarray}
z^{2} J^{2} + \gamma^{2} &=& \left( z^{4} J^{4} + 4 \gamma^{4} \right) T_{0i}^{\uparrow} T_{0j}^{\uparrow} + 2 z^{2} J^{2} \gamma^{2} \Big[ (T_{0i}^{\uparrow})^{2} \nonumber \\ 
&+& (T_{0j}^{\uparrow})^{2} \Big] + \gamma \left( z^{2} J^{2} - 2 \gamma^{2} \right) \left( T_{0i}^{\uparrow} - T_{0j}^{\uparrow} \right).~~~~~~~~
\label{cric_J_cb}
\end{eqnarray}
Here, $z$ is the coordination number and it is $4$ for the present work. The sublattice coefficients $T_{0i}^{\sigma}$ and $T_{0j}^{\sigma}$ in the above equation, representing the effective hopping contributions are
\begin{subequations}
\begin{eqnarray}
T_{0i}^{\sigma} &=& \frac{1}{J_{0ci}^{\sigma}} = \frac{n_i^{\sigma}+1}{\mathcal{U}_{i} + \mathcal{V}_{i\theta} - \mu} 
                                                -\frac{n_i^{\sigma}}{\mathcal{U}_{i} - U + \mathcal{V}_{i\theta} - \mu}, ~~~~~~~~~\\
T_{0j}^{\sigma} &=& \frac{1}{J_{0cj}^{\sigma}} = \frac{n_j^{\sigma}+1}{\mathcal{U}_{j} + \mathcal{V}_{j\theta} - \mu} 
                                                -\frac{n_j^{\sigma}}{\mathcal{U}_{j} - U + \mathcal{V}_{j\theta} - \mu},~~~~~~~~~
\end{eqnarray}
\label{eff_hop}
\end{subequations}
where $\mathcal{U}_{i} = U n_i^{\sigma} + U_{\uparrow\downarrow} n_i^{\sigma'}$ and $\mathcal{V}_{i\theta} = 2 V n_j^{\sigma} + 2 V(\theta) n_j^{\sigma} + 2 V_{\uparrow \downarrow} n_{j}^{\sigma'} + 2 V_{\uparrow \downarrow}(\theta) n_{j}^{\sigma'}$. The 
$\mathcal{U}_{j}$ and $\mathcal{V}_{j\theta}$ are obtained by replacing the sublattice lattice indices $i$ by $j$. 

Following similar steps, we further discuss the perturbative analysis for phase transitions corresponding to stripe-ordered phases at and above the magic angles. The unperturbed ground-state energy at sublattices $i$ and $j$ is the same as in Eq.~(\ref{ener_up}). The perturbed Hamiltonian $\hat{H}_{per}$ for stripe density wave (SDW) to compressible stripe supersolid (or SF) phase transitions is expressed as
\begin{equation}
\begin{aligned}
\hat{H}^{s}_{per} = &-J \sum_{\sigma} \Big[ \bar{\phi}_i^{\sigma x} (\hat{b}_i^{\sigma \dagger} + \hat{b}_i^{\sigma}) + \bar{\phi}_j^{\sigma x} (\hat{b}_j^{\sigma \dagger} + \hat{b}_j^{\sigma}) \\
&\quad + \bar{\phi}_i^{\sigma y} (\hat{b}_i^{\sigma \dagger} + \hat{b}_i^{\sigma}) + \bar{\phi}_j^{\sigma y} (\hat{b}_j^{\sigma \dagger} + \hat{b}_j^{\sigma}) \Big] \\
&+ \gamma_x \Big[ \bar{\phi}_i^{x \downarrow} (\hat{b}_{i}^{\uparrow \dagger} - \hat{b}_{i}^{\uparrow}) + \bar{\phi}_i^{x\uparrow} (\hat{b}_{i}^{\downarrow \dagger} - \hat{b}_{i}^{\downarrow}) \Big] \\
&+ i\gamma_y \Big[ \bar{\phi}_i^{y \downarrow} (\hat{b}_{i}^{\uparrow \dagger} - \hat{b}_{i}^{\uparrow}) + \bar{\phi}_i^{y \uparrow} (\hat{b}_{i}^{\downarrow \dagger} - \hat{b}_{i}^{\downarrow}) \Big] \\
&+ \gamma_x \Big[ \bar{\phi}_j^{x \downarrow} (\hat{b}_{j}^{ \uparrow \dagger} - \hat{b}_{j}^{\uparrow}) + \bar{\phi}_j^{x\uparrow} (\hat{b}_{j}^{\downarrow \dagger} - \hat{b}_{j}^{\downarrow}) \Big] \\
&+ i\gamma_y \Big[ \bar{\phi}_j^{y \downarrow} (\hat{b}_{j}^{\uparrow \dagger} - \hat{b}_{j}^{\uparrow}) + \bar{\phi}_j^{y \uparrow} (\hat{b}_{j}^{ \downarrow \dagger} - \hat{b}_{j}^{\downarrow}) \Big],
\end{aligned}
\label{ham_pert_stripe}
\end{equation}
where $\bar{\phi}_{i}^{\sigma x} = \sum_{\langle j \rangle i} \phi_{j}^{\sigma x}$ and $\bar{\phi}_{i}^{\sigma y} = \sum_{\langle j \rangle i} \phi_{i}^{\sigma y}$ are the sum of nearest-neighbour order parameters along the $x$- and $y$-directions, respectively. The leading second-order correction in the energy and its minimization are discussed in the appendix~\ref{sec-II}. The phase boundary of SDW-SF(SSS) phase transitions is obtained by
\begin{equation}
  \left (\frac{z^2J^2}{4} + \gamma^{2} \right )(T_{0i}^{\uparrow} + T_{0j}^{\uparrow}) = \frac{zJ}{2},
  \label{pb_str}
\end{equation}
where the effective hoppings are defined in Eq.~(\ref{eff_hop}).

Finally, the presence of finite interspin interaction leads to correlated quantum phases. The variation of tilt angle further gives rise to transitions between ordered correlated phases. To discuss the phase boundary for below the magic angle cases with checkerboard structure, the $V_{\uparrow\downarrow}$ term is retained 
$(\zeta\neq 0)$ in the ground state energy [Eq.~(\ref{ener_up})]. The perturbed Hamiltonian is given by Eq.~(\ref{ham_per}). At the cMI--SF(SS) or cDW--SF(SS)
phase boundary, the effective hopping parameters satisfy $T_{0i}^{\uparrow} = T_{0i}^{\downarrow}$ and $T_{0j}^{\uparrow} = T_{0j}^{\downarrow}$, which leads to a pairwise degeneracy of the eigenvalues. The minimization of second-order correction in energy gives the boundary of cDW-SF(SS) and cMI-SF(SS) phase transitions
\begin{equation}
   z^{2} J^{2} + \gamma^{2} = \left( z^{2} J^{2} + 2\gamma^{2} \right)^{2} T_{0i}^{\uparrow} T_{0j}^{\uparrow}.
   \label{pb_cdw}
\end{equation}
The above equation characterizes various incompressible - compressible phase transitions of correlated phases in the presence of spin-orbit coupling. Moreover, for the 
transitions above the magic angle where various correlated stripe-ordered quantum phases appear, the phase boundary can be obtained with the same sublattice occupancy
as done for below magic angles. This leads to the phase boundary as obtained in Eq.~(\ref{pb_str}) but now with finite $V_{\uparrow\downarrow}$. The predictions of mean-decoupling analysis for the critical hopping amplitude of various phase boundaries of staggered and correlated quantum phases are studied with numerical results.   
\section{Ground State Phase Diagrams}
\label{res}
We present equilibrium phase diagrams of spin-orbit-coupled dipolar bosons in a square lattice. The dipoles are polarized in the $yz$-plane and the tilt of polarization axis of the dipoles makes an angle $\theta$ with the $z$-axis. We obtain the phase diagrams at different SOC strengths to characterize the novel staggered phase due to interplay of dipolar and spin-orbit interaction. To obtain the phase diagrams, we use the initial state with random distribution of Gutzwiller coefficients across the lattice. This choice gives corresponding initial SF order parameter with random amplitude and phase. The Fock space dimension at each lattice site is $n_{\rm max} = 6$. We considered $20$ random configurations of the initial states to ensure the ground state of the system. We set $\mu/J = 10$ and $U_{\uparrow\downarrow}/U = 0.5$. We first examine the effects of intraspin offsite interactions and then include the role of interspin NN interaction in the emergence of correlated finite-momentum superfluids and supersolids.  

\subsection{Intraspin dipolar interactions}
\subsubsection{$\theta = 0^\circ$ and $\theta < \theta_{M}$}
\begin{figure}[h]
    \includegraphics[width=\linewidth]{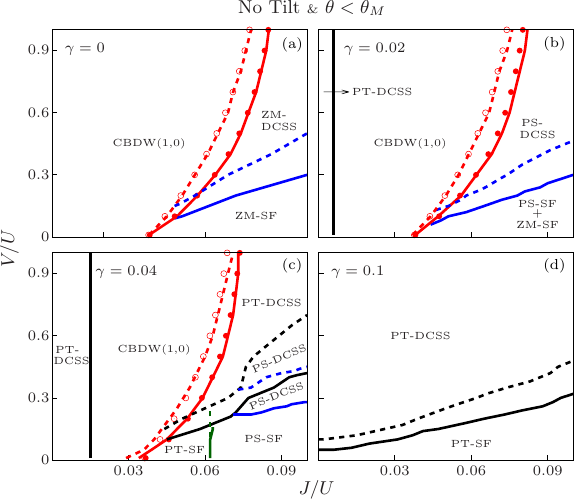}
    \caption{Ground-state phase diagrams for SOC strengths (a) $\gamma = 0$, (b) $\gamma = 0.02$, (c) $\gamma = 0.04$ and $\gamma = 0.1$ with $\zeta=0$. The solid (dashed) lines are the phase boundaries for the tilt angle $\theta = 0^{\degree}$ ($30^{\degree}$) in the hopping amplitude and dipolar interaction strength plane. The red filled (open) circles are the phase boundaries between incompressible and compressible phases, obtained by mean-field decoupling theory for $\theta = 0^{\degree}$ ($30^{\degree}$) [Eq.~(\ref{cric_J_cb})]. Here, $U_{\uparrow\downarrow}/U = 0.5$ and $\mu/J = 10$.}
    \label{pd_0to30}
\end{figure}

Fig~\ref{pd_0to30} presents the phase diagrams for a range of spin-orbit coupling strengths $\gamma \in \{0, 0.02, 0.04, 0.1\}$, focusing on the tilt angles below the magic angle ($\theta < 35.3^\circ$). In this regime, the dipolar interaction is characterized by a repulsive nature along both $x$- and $y$-directions. The phase boundaries for the isotropic case ($\theta = 0^\circ$) are indicated by solid lines, while those for the anisotropic case ($\theta = 30^\circ$) are shown as dashed lines. Furthermore, the numerical incompressible-compressible phase boundaries are corroborated by analytical results derived from Eq.~(\ref{cric_J_cb}), which are marked by filled red and open circles for the respective tilt angles.
\begin{figure}[h]
    \includegraphics[width=\linewidth]{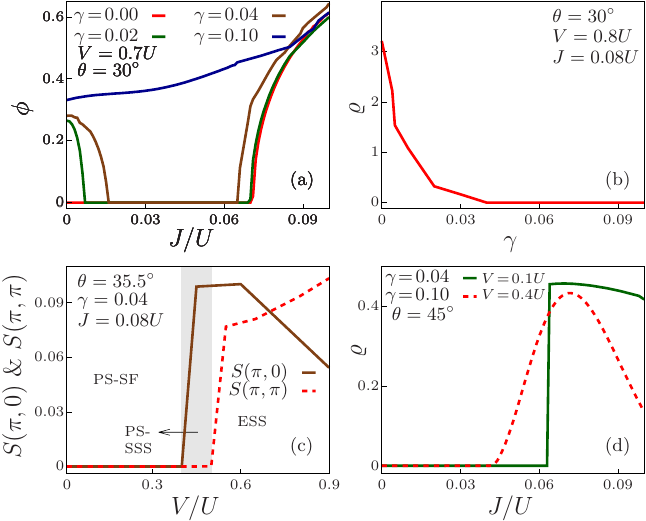}
    \caption{(a) The average SF order parameter as a function of hopping amplitude at $V=0.7$ and $\theta = 30^\circ$ for different SOC    
    strengths, that demarcates density-modulating compressible phases on either sides of CBDW(1,0) phase. (b) The spin-dependent momentum measure characterizing PT-DCSS to PS-DCSS transition with $\gamma$. (c) Structure factor at $(\pi,\pi)$ and $(\pi,0)$ distinguishing different orders of magnetic supersolid phases at $J=0.08$ and $\theta = 35.5^\circ$. (d) $\varrho$ with hopping amplitude for two cases corresponding to PT-SF to PS-SF  (green solid line) and PT-SSS to PS-SSS (red dashed line) phase transition.}
    \label{op_variation}
\end{figure}
As illustrated in Fig.~\ref{pd_0to30}(a), the system corresponds to the two-component dipolar bosons that possess isotropic NN interactions ($\theta = 0^\circ$) with $\gamma = 0$. By tuning the kinetic energy via the hopping amplitude $J/U$, the system undergoes a quantum phase transition from the incompressible checkerboard density-wave phase to compressible phases. For smaller values of $V/U$, the system remains either in the DW or SF phase. Among these, the SF phase does not exhibit diagonal long-range order, whereas the DW phase lacks ODLRO and possesses only diagonal order. In contrast, for larger values of $V/U$, phases with diagonal order become energetically favorable. In this regime, the DCSS phase emerges, characterized by the coexistence of diagonal order and ODLRO. Notably, at $\theta = 30^\circ$, the incompressible phase region undergoes a significant contraction. This reduction in the region of CBDW phase is primarily attributed to the anisotropic nature of the dipolar interactions; specifically, the interaction strength along the $y$-axis diminishes as a function of the tilt angle $\theta$, thereby destabilizing the crystalline order of the CBDW state. This is consistent to the previous studies on single-component dipolar bosons~\cite{zhang_21,bandyopadhyay_19}.

The introduction of a finite spin-orbit coupling significantly modifies the phase landscape. The SOC induces the formation of distinct spatial textures, specifically phase-twisted and phase-striped structures. This leads to the emergence of finite-momentum superfluid and supersolid states namely PT-DCSS, PS-DCSS, PT-SF, and PS-SF, respectively. As synthetic SOC is a complex spin-dependent hopping, it enhances the effective kinetic energy of the particles. Consequently, an increase in $\gamma$ facilitates the transition from incompressible density-ordered phases to compressible regimes, resulting in shrinking of the insulating phase~\cite{yan_17,suthar_21,pu_24}. As shown in Fig.~\ref{pd_0to30}(b), at $\gamma = 0.02$, the CBDW region is destroyed at smaller $J$, and a compressible phase with finite $S(\pi,\pi)$ and zero $\varrho$, identified as PT-DCSS emerges. The CBDW-DCSS(or SF) phase boundary shifts towards lower $J$ with finite 
$\gamma$. At higher $J$, the zero-momentum SF and SS regime are occupied by the PS-SF and PS-DCSS states at the weak and strong dipolar interaction strengths, respectively. We find that some traces of ZM-SF phase still remain at $\gamma=0.02$ for higher $J$'s. At both tilt angles, the incompressible-compressible phase boundary agree well with the analytical perturbative analysis results (filled and open circles).

With an increase in $\gamma$ at $\gamma = 0.04$ [Fig.~\ref{pd_0to30}(c)], the phase diagram remains qualitatively similar. However, the incompressible phase region continues to diminish, by supporting finite-$\mathbf{k}$ compressible phases. In addition, near the insulating domain, the twisted SF phase dominates and thus reduces the PS-DCSS and PS-SF phase region. Fig.~\ref{op_variation}(a) presents the variation of SF order parameter, which confirms the shift in PT-DCSS phase boundary towards higher $J$'s, leading to narrowing insulating domains. Furthermore, the existence of the twisted phase with $\gamma$ is verified by the change in the spin-dependent momentum $\varrho$ from finite to zero [Fig.~\ref{op_variation}(b)]. Moreover, the same order parameter also characterizes the PT-SF to PS-SF phase transition at lower $V$'s. Upon reaching a higher coupling strength of $\gamma = 0.1$ [Fig.~\ref{pd_0to30}(d)], the incompressible phase vanishes entirely from the phase diagram. In this strong-SOC limit, the system is dominated by the PT-DCSS and PT-SF phases, indicating that the enhanced spin-dependent hopping effectively destabilizes the density-wave phase. Thus, the SF order parameter remains finite as a function of hopping amplitude, characterizing finite-momentum compressible states [cf.Fig.~\ref{op_variation}(a)]. 

\subsubsection{$\theta\approx\theta_{M}$ (close to magic angle)}
At the magic angle $\theta = \theta_M \approx 35.3^\circ$, the dipolar interaction along the $y$-direction is completely suppressed, while the interaction along the $x$-direction remains finite and repulsive. This pronounced anisotropy of the interaction leads to an imbalance in the effective interaction energy, favoring atomic redistribution along the direction of vanishing interaction. Consequently, the system stabilizes striped density-wave configurations rather than checkerboard or diagonal orderings, as such stripe patterns minimize the total dipolar interaction energy.
\begin{figure}[ht]
    \includegraphics[width=\linewidth]{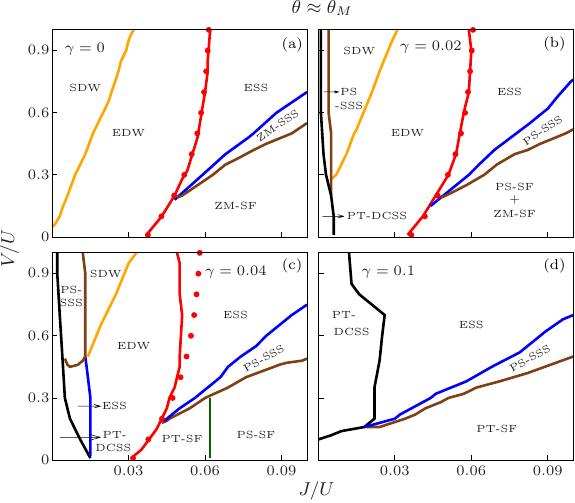}
    \caption{Ground-state phase diagrams for $U_{\uparrow\downarrow}/U = 0.5$ and $\mu/J = 10$ with SOC strengths (a) $\gamma = 0$, (b) $0.02$, (c) $0.04$, and (d) $0.1$. The tilt angle of polarized dipolar bosons is $\theta = 35.5^{\degree}$ with $\zeta = 0$. Phase boundaries marked by red filled circles are obtained from analytical mean-field decoupling theory [Eq.~(\ref{pb_str})].}
    \label{pd_35}
\end{figure}     
The phase diagrams close to the magic angle for different SOC strengths are presented in Fig.~\ref{pd_35}. As depicted in Fig.~\ref{pd_35}(a), for $\theta = 35.5^\circ$ and $\gamma = 0$, the dipolar interaction along the $y$-direction becomes attractive. In the limit of weak hopping, this attraction stabilizes two insulating phases, namely the stripe density wave and the emulsion density wave (EDW). With increasing hopping strength $J/U$, the system undergoes a transition to compressible phases. At lower dipolar interaction strengths, a ZM-SF phase appears. For stronger interactions, the competition between the kinetic energy and attractive dipolar interactions gives rise to the stripe supersolid (SSS) and emulsion supersolid (ESS) phases. It is important to note that for the emulsion states, the structure factor $S(\mathbf{k})$ remains finite at both $(\pi,\pi)$ and $(\pi,0)$, and thus, can be distinguished from the phases ordered in stripes.

In Fig~\ref{pd_35}(b) for finite SOC strength at $\gamma = 0.02$, the system exhibits PT and PS structures in the supersolid phase. Due to anisotropy along $y$- direction, we get stripe phase i.e SSS and ESS. Furthermore, with increase in the SOC strengths, the region of incompressible phases (EDW \& SDW) decrease and compressible phases increase. At lower hopping strengths, the PS-SSS phase is stabilized by the dipolar interaction. Upon increase in hopping strengths, we get incompressible SDW and EDW phases. And, further increasing the value of hopping leads to superfluid states at lower $V$'s and supersolid phases at higher $V$'s. The intervening incompressible phases arise due to repulsive NN interaction along $x$-direction which is always repulsive.

With a further increase in the SOC strength to $\gamma = 0.04$, the incompressible region of the phase diagram shrinks, while the compressible region expands. This is expected as SOC promotes finite-momentum superfluidity by melting the insulating states. In the low-hopping regime, various supersolid phases such as PT-DCSS, PS-SSS, and ESS are stabilized. As the hopping amplitude increases, the system transitions into density-wave phases, namely SDW and EDW. Upon further increasing the hopping strength, the system enters the compressible SS and SF phases. At lower values of $V$, the PT-SF and PS-SF phases are observed, whereas at higher $V$, the PS-SSS and ESS phases become energetically favorable. At larger hopping strength, the transition from PS-SF to ESS, through PS-SSS can be traced by the structure factor, as shown in Fig.~\ref{op_variation}(c) for $J/U = 0.08$. Note that the numerical phase boundaries between incompressible EDW and compressible phases are in good agreement with the analytical mean-field decoupling theory (red filled circles) for all three values of SOC strengths [Fig.~\ref{pd_35}(a,b,c)]. 

For $\gamma = 0.1$, the incompressible phase completely disappears, and the system remains in compressible phases throughout the $J$-$V$ plane. This is evident from Fig.~\ref{pd_35}(d). In the regime of weaker dipolar interaction, the PT-SF phase is stabilized. As $V$ increases, the system transitions to PT-DCSS, PS-SSS, and ESS phases, which dominate the higher interaction regime.

\subsubsection{$\theta > \theta_{M}$}
For tilt angles exceeding the magic angle, the dipolar interaction along the $y$-direction becomes attractive. This attractive coupling further promotes the formation of stripe-ordered phases by energetically favoring particle alignment along the transverse direction. As a representative case, we consider $\theta = 45^\circ$. At this angle, the interaction along the $y$-direction becomes strongly attractive, which leads to the stabilization of the SDW and SSS phases.
\begin{figure}[ht]
    \includegraphics[width=\linewidth]{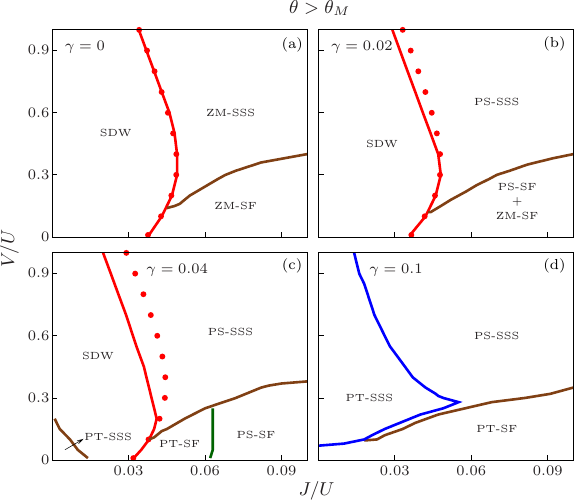}
    \caption{Ground-state phase diagrams for $U_{\uparrow\downarrow}/U = 0.5$ and $\mu/J = 10$ with SOC strengths (a) $\gamma = 0$, (b) $0.02$, (c) $0.04$ and  (d) $0.1$. The tilt angle is $\theta = 45^{\degree}$ and $\zeta = 0$. Phase boundaries marked by red filled circles are obtained from analytical theory [Eq.~(\ref{pb_str})].}
    \label{pd_45}
\end{figure}

In Fig.~\ref{pd_45}(a), for $\gamma = 0$, the system resides in the SDW phase in the low-hopping regime. As the hopping amplitude $J/U$ increases, the system undergoes a transition from the incompressible SDW phase to compressible phases. For smaller values of $V$, the compressible phase corresponds to the ZM-SF state, whereas for larger $V$ it evolves into the ZM-SSS phase.

For a finite spin-orbit coupling, $\gamma = 0.02$ [Fig.~\ref{pd_45}(b)], the extent of the SDW phase is reduced, while the compressible phase region expands. In this regime, both PS-SF and ZM-SF phases appear at lower dipolar interactions, whereas the PS-SSS phase stabilized at higher $V$ values. With further increase in SOC strength to $\gamma = 0.04$ [Fig.~\ref{pd_45}(c)], the incompressible region shrinks further and the compressible region correspondingly enlarges. Additionally, a PT-SSS phase emerges in the lower-left corner of the phase diagram, along with 
PT-SF and PS-SF phases at lower $V$, and the PS-SSS phase at higher $V$. The transition to PS-SF phase (at higher $J$) is characterized by the variation of $\varrho$ with $J/U$, as shown in Fig.~\ref{op_variation}(d) at $V=0.1$. Finally, for strong SOC strengths [Fig.~\ref{pd_45}(d)], $\gamma = 0.1$, the incompressible phase completely disappears, and only compressible phases remain, namely PT-SSS, PS-SSS, and PT-SF, similar to Fig.~\ref{pd_0to30}(d) and Fig.~\ref{pd_35}(d). The unique spin-dependent momentum order parameter $\varrho$ varies from zero to finite at PT-SSS to PS-SSS phase transition. It is plotted for $V=0.4$ in Fig.~\ref{op_variation}(d). Thus, the interplay of anisotropic dipolar (minimal NN) interaction and spin-orbit coupling leads to novel staggered finite-momentum superfluid and supersolid phases. 

\subsection{Interspin dipolar interactions and correlated phases}
\subsubsection{$\theta = 0^\circ$ and $\theta < \theta_{M}$}
\begin{figure}[h]
    \includegraphics[width=\linewidth]{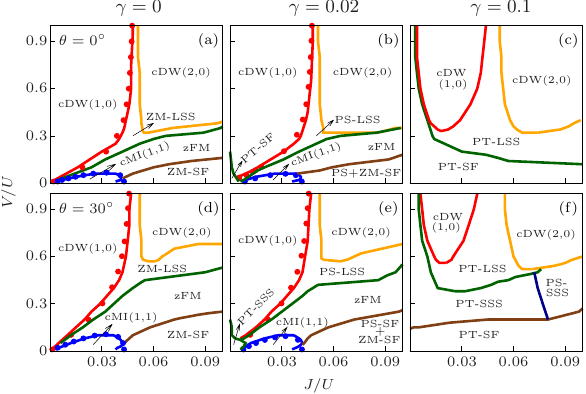}
    \caption{Ground-state phase diagrams in presence of interspin NN interactions $V_{\uparrow\downarrow}=\zeta V$ where $\zeta = 1$ for $U_{\uparrow\downarrow}/U = 0.5$ and $\mu/J = 10$ with SOC strengths $\gamma = 0$, $0.02$, and $0.1$ in (a,d), (b,e), and (c,f). The tilt angle is $\theta = 0^{\degree}$ for upper panel (a,b,c) and $30^{\degree}$ for lower panel (d,e,f). The phase boundaries marked by the red and blue filled circles are obtained from mean-field decoupling theory [Eq.~(\ref{pb_cdw})].}
    \label{interpd_0_30}
\end{figure}

Fig~\ref{interpd_0_30} represents phase diagrams at tilt angle $0^\circ$ and $30^\circ$ degree for soc strengths $\gamma \in \{0, 0.02, 0.1\}$ in the presence of interspin dipolar interactions $V_{\uparrow \downarrow} = V$ i.e $\zeta = 1$. In fig~\ref{interpd_0_30}(a) for $\gamma = 0$ and $\theta = 0^\circ$ we get two incompressible phases i.e the correlated MI (cMI) and correlated DW (cDW) phases. The system exhibits non-trivial density distribution characteristics in which the total atomic occupancy at each lattice site remains an integer, while the number of atoms belonging to each spin component fluctuates randomly. Therefore, the cMI and cDW phases are characterized by the pair $(N_i, N_j)$.
In the $\text{cDW}(1,0)$ phase, the occupations satisfy $N_j = 0$ and $N_i = 1$. However, the individual spin occupations $n_i^{\uparrow}$ and $n_i^{\downarrow}$ at $i$th site are randomly distributed between $0$ and $1$, while still satisfying the constraint of a fixed total occupation. For $\text{cMI}(1,1)$ phase, the total densities $N_i = N_j =1$. And, the spin occupancy satisfy the condition $n_{i}^{\uparrow} = 1 - n_{i}^{\downarrow}$ with $n_{i}^{\downarrow} \in \{0,1\}$, where the values between the two possibilities are chosen at random. The same distribution will hold for $j$ sub-lattice. As hopping increases we reveal a transition from incompressible to compressible phase at lower dipolar interactions, i.e lattice supersolid (LSS), $z$-polarized ferromagnetic (zFM) and ZM-SF compressible phases. The LSS phase is characterized by real occupation numbers $n_i^{\sigma}$ with $n_i^{\sigma} \neq n_j^{\sigma}$, and the density imbalance $\Delta n^{\sigma}$ being real [see Fig.~\ref{phase_lss}]. In the zFM phase, the SF order parameter of only one of the component acquires a finite value, while that of other remains zero. Correspondingly, the occupation number density of only one spin component remains finite. For instance, if $n_i^{\uparrow}$ is finite, then $n_i^{\downarrow}=0$, and vice-versa. At small hopping amplitudes and weak dipolar interactions, the system resides in the incompressible cMI and cDW phases. As the hopping strength increases, these insulating phases gradually disappear and the system transitions into compressible phases. The corresponding phase boundaries agree with the analytical mean-field decoupling results, shown by (red and blue) filled circles. For larger values of the dipolar interaction, the ZM-LSS phase appears in a narrow region sandwiched between two incompressible density-wave phases, namely 
$\mathrm{cDW}(1,0)$ and $\mathrm{cDW}(2,0)$. 

As the strength of SOC increases, as shown in Fig.~\ref{interpd_0_30}(b) for $\gamma = 0.02$, a compressible PT-SF phase appears in the region of low hopping and weak dipolar interaction. With further increase in the hopping amplitude at low dipolar interaction strength, a coexistence of PS-SF and ZM-SF phases is observed. In addition, the presence of SOC induces the PS-LSS phase [see Fig.~\ref{phase_lss}], which emerges between two incompressible phases. Here, the phase $\theta^{\sigma}$ possesses a stripe variation throughout the lattice, distinguishing it from its zero-momentum counterpart. It is also noticeable that the region corresponding to the incompressible phases shrinks with increasing SOC strength, while the overall structure of the remaining phases largely remains similar to Fig.~\ref{interpd_0_30}(a). With a further increase in SOC $(\gamma = 0.1)$, we observe PT-SF and PT-LSS throughout the hopping axis for lower values of dipolar interaction $V/U$ [Fig~\ref{interpd_0_30}(c)]. And, at higher value of $V/U$ the compressible phase PT-LSS intervenes the incompressible phases $\text{cDW(1,0)}$ and $\text{cDW(2,0)}$.
\begin{figure}[h]
    \includegraphics[width=\linewidth]{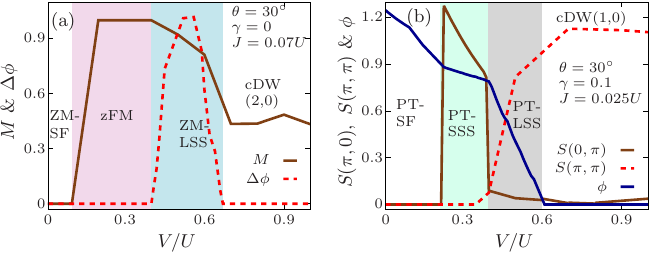}
    \caption{(a) Variation of the magnetization $\mathcal{M}$ and SF-order parameter imbalance $\Delta \phi$ as a function of the dipolar interaction $V/U$ for $\theta = 30^\circ$, $\gamma = 0$, and $J = 0.07U$. Different quantum phases such as ZM-SF, zFM, ZM-LSS, and cDW$(2,0)$ are identified, with phase boundaries indicated by shaded regions. (b) Corresponding behavior of the structure factors $S(0,\pi)$, $S(\pi,\pi)$, and SF-order parameter $\phi$ as a function of $V/U$ for $\theta = 30^\circ$, $\gamma = 0.1$, and $J = 0.025U$. The emergence of PT-SF, PT-SSS, PT-LSS, and cDW$(1,0)$ phases is highlighted.}
    \label{inter_order_parameters}
\end{figure}

For $\theta = 30^\circ$ and $\gamma = 0$, as shown in Fig.~\ref{interpd_0_30}(d), the effective dipolar interaction strength decreases (in $y$ direction). As a result, the region of the incompressible phase along $V/U$ becomes smaller compared to Fig.~\ref{interpd_0_30}(a). On the other hand, the compressible phase region expands along the $J/U$ direction. To distinguish the zFM phase, we define the magnetization $\mathcal{M} = {|(n_{i}^{\uparrow} - n_{j}^{\downarrow})|}/(n_{i}^{\uparrow} + n_{j}^{\downarrow})$. Fig.~\ref{inter_order_parameters}(a) shows the variation of magnetization $\mathcal{M}$ and the SF order parameter imbalance $\Delta \phi$ as a function of $V$ at $J/U = 0.07$. $\mathcal{M}$ takes finite value as system enters (from ZM-SF phase) to zFM phase and further decreases and stabilizes in correlated LSS and cDW phases. For a finite SOC strength $(\gamma = 0.02)$, as shown in Fig.~\ref{interpd_0_30}(e), a compressible PT-SSS phase appear at lower values of hopping and dipolar interaction. For $V/U \gtrsim 0.6$, a compressible phase (PS-LSS) is observed, and at higher $V$ values, it is sandwiched between two incompressible cDW phases. With a further increase in SOC strength $(\gamma = 0.1)$, as illustrated in Fig.~\ref{interpd_0_30}(f), the region of the compressible phase expands. For lower dipolar interaction strengths, PT-SF, PT-SSS, and PS-SSS phases are observed across the entire hopping axis. In contrast, at higher dipolar interaction strengths, a PT-LSS phase appears in between the incompressible phases \text{cDW(1,0)} and \text{cDW(2,0)}. The variation of the structure factors $S(0,\pi)$ and $S(\pi,\pi)$, along with the SF order parameter $\phi$, illustrating the transitions between PT-SF, PT-SSS, PT-LSS, and cDW phases in the presence of SOC, as presented in Fig.~\ref{inter_order_parameters}(b).

\subsubsection{$\theta \gtrapprox \theta_{M}$}
\begin{figure}[h]
    \includegraphics[width=\linewidth]{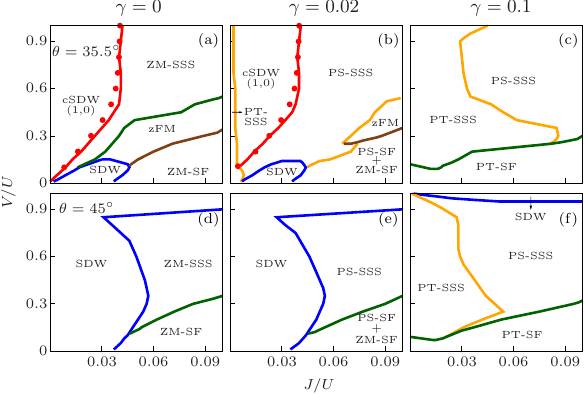}
    \caption{Ground-state phase diagrams in presence of interspin NN interactions $V_{\uparrow\downarrow}=\zeta V$ where $\zeta = 1$ for $U_{12}/U = 0.5$ and $\mu/J = 10$ with SOC strengths $\gamma = 0$, $0.02$ and $0.1$ in (a) \& (d), (b) \& (e) and (c) \& (f). The angle is $\theta = 35.5^{\degree}$ from (a)-(c) and $45^{\degree}$ from (d)-(f). The cSDW phase possesses the stripe-order in total occupancy while SDW phase has stripe-ordered variation in individual components. Phase boundaries marked by red filled circles are obtained from 
    mean-field decoupling theory [Eq.~(\ref{pb_str})].}
    \label{interpd_35p5_45}
\end{figure}

We finally discuss the correlated quantum phases near and above the magic tilt angle of dipolar bosons. For an angle close to the magic angle, $\theta = 35.5^\circ$ and in the absence of spin-orbit coupling ($\gamma = 0$), as shown in Fig.~\ref{interpd_35p5_45}(a), the system exhibits incompressible phases such as correlated stripe-density wave (cSDW) and stripe-density wave at low hopping amplitudes. In the cSDW$(1,0)$ phase, the total occupation numbers satisfy $N_i=1$ and $N_j=0$ along one of the directions. However, within the occupied sites, the spin-resolved densities $n_i^{\uparrow}$ and $n_i^{\downarrow}$ are randomly distributed between $0$ and $1$ along the stripe pattern. With an increase in hopping strength we reveal a transition from incompressible to compressible phases i.e ZM-SF, zFM, \& ZM-SSS. For a finite spin-orbit coupling $(\gamma = 0.02)$, the region corresponding to the incompressible phases shrinks, while the compressible phases expand [Fig.~\ref{interpd_35p5_45}(b)]. The incompressible-compressible phase boundaries agree with the 
mean-field decoupling theory discussed in Sec.~\ref{mfdt}.
At low values of both hopping and dipolar interaction strength, the system exhibits a PT-SSS phase. As the hopping amplitude increases, a coexistence region of PS-SF and ZM-SF phases appears. Furthermore, with increasing dipolar interaction strength, the system transitions into the zFM and PS-SSS phases. With a further increase in the spin-orbit coupling strength $(\gamma = 0.1)$, as illustrated in Fig.~\ref{interpd_35p5_45}(c), the incompressible phases completely disappear, and the phase diagram is dominated by compressible finite-momentum phases. At lower values of the dipolar interaction strength, the system exhibits a PT-SF phase. As the dipolar interaction increases, the system transitions into density-modulated PT-SSS and PS-SSS phases.

For $\theta = 45^\circ$ and in the absence of spin--orbit coupling $(\gamma = 0)$, as shown in Fig.~\ref{interpd_35p5_45}(d), the system exhibits a transition from the SDW phase to the ZM-SF phase at lower values of dipolar interaction as the hopping amplitude increases. At higher dipolar interaction strengths, the SDW phase instead transitions into a ZM-SSS phase. Furthermore, for $V/U \geq 0.9$, the incompressible phase persists throughout the range of hopping, thereby dominating the phase diagram. Here, the stripe order phases appear with multiple sublattice occupancies, thus we refrain from obtaining the analytical phase boundary. At finite spin-orbit coupling $(\gamma = 0.02)$, as shown in Fig.~\ref{interpd_35p5_45}(e), the ZM-SSS phase observed in Fig.~\ref{interpd_35p5_45}(d) transforms into a corresponding PS-SSS phase due to the presence of SOC. Similarly, the ZM-SF phase evolves into a region where both PS-SF and ZM-SF phases coexist
[cf. Fig.~\ref{interpd_35p5_45}(d) and Fig.~\ref{interpd_35p5_45}(e)]. With a further increase in the spin-orbit coupling strength $(\gamma = 0.1)$, as shown in Fig.~\ref{interpd_35p5_45}(f), the phase diagram is predominantly occupied by compressible phases. This is attributed to 
the role of complex hopping in melting the incompressible states, at the cost of emerging finite-$k$ phases. At lower values of the dipolar interaction strength, the system exhibits a PT-SF phase. As the dipolar interaction increases, the system transitions into PT-SSS and PS-SSS phases. However, for $V/U \geq 0.95$, an incompressible SDW phase reappears in the phase diagram.

\section{Conclusions}
\label{conc}
We have studied the ground-state phases of two-component dipolar bosons in a square lattice in the presence of synthetic spin-orbit coupling and anisotropic nearest-neighbour dipolar interactions. Using the site-decoupled Gutzwiller mean-field approach and analytical mean-field decoupling theory, we obtained the phase diagrams for different tilt angles of magnetic dipoles and SOC strengths. To characterize various finite-momentum and correlated quantum phases, we used spin-dependent momentum distribution, phase angle ordering, structure factors, and magnetization. Our results reveal the intriguing role of the tilt of the dipole moments in determining the nature of the ordered quantum phases. For tilt angles close to and greater than the magic angle, the anisotropic dipolar interaction along the $y$-direction becomes attractive, leading to the emergence of stripe-like density-wave configurations such as SDW and correlated SDW phases. As the hopping strength increases, these insulating phases gradually transition to compressible phases, including superfluid and supersolid states. We further find that the presence of spin-orbit coupling remarkably modifies the phase angle distribution of the superfluid order parameter. The spin-orbit coupling reduces the region of incompressible density-ordered phases while expanding the domain of compressible phases such as phase-twisted and phase-stripe superfluids and supersolids. In the strong SOC regime, the insulating phases are largely suppressed, and the system is dominated by finite-momentum superfluid and supersolid states. Our study reveals a rich interplay between spin-orbit coupling, dipolar interaction anisotropy, and hopping amplitude, leading to varieties of exotic quantum phases. These findings provide a way to realize novel supersolid states of spin-orbit coupled dipolar bosons in optical lattice experiments.

\begin{acknowledgments}
N.K. and K.S. acknowledge support from the Science and Engineering Research Board, Department of Science and Technology, Government of India through Project No. SRG/2023/001569.
\end{acknowledgments}
\appendix
\section{CBDW-SF(SS) phase boundary}
\label{sec-I}
For the canted spin-orbit coupled dipolar bosons, the single-particle hopping and spin-orbit coupling (complex spin-dependent hopping) terms of the single-site Hamiltonian are treated as perturbation while the on-site and off-site NN interaction terms along with the chemical potential serve as unperturbed Hamiltonian. The energy of the ground states corresponding to an unperturbed Hamiltonian is given as Eq.~(\ref{ener_up}) and the perturbed Hamiltonian is defined in Eq.~(\ref{ham_per}). Using perturbation theory, the second-order perturbed ground state energy is 
\begin{eqnarray}
 E^{(2)}_{n_i^\uparrow, n_i^\downarrow; \, n_j^\uparrow, n_j^\downarrow}
&=& \sum_{m \neq n}
\frac{
\left|
\left\langle 
m_i^\uparrow, m_i^\downarrow, m_j^\uparrow, m_j^\downarrow
\middle| 
\hat{H}^{c}_{per}
\middle|
n_i^\uparrow, n_i^\downarrow, n_j^\uparrow, n_j^\downarrow
\right\rangle
\right|^2
}{
E^{(0)}_{n_i^\uparrow, n_i^\downarrow} - E^{(0)}_{m_i^\uparrow, m_i^\downarrow}
} \nonumber \\
&=& z^{2} J^{2} \Big[ |\phi_{i}^{\uparrow}|^{2} T_{0j}^{\uparrow} + |\phi_{i}^{\downarrow}|^{2} T_{0j}^{\downarrow} + |\phi_{j}^{\uparrow}|^{2} T_{0i}^{\uparrow} + |\phi_{j}^{\downarrow}|^{2} T_{0i}^{\downarrow} \Big] \nonumber \\
&+& 2 z J \Big( \phi_{i}^{\uparrow}\phi_{j}^{\uparrow} + \phi_{i}^{\downarrow}\phi_{j}^{\downarrow} \Big) + 2 \gamma \Big( \phi_{i}^{\uparrow}\phi_{j}^{\downarrow} - \phi_{i}^{\downarrow}\phi_{j}^{\uparrow} \Big) \nonumber \\
&+& 2 \gamma^{2} \Big[ |\phi_{i}^{\uparrow}|^{2} T_{0j}^{\downarrow} + |\phi_{i}^{\downarrow}|^{2} T_{0j}^{\uparrow} + |\phi_{j}^{\uparrow}|^{2} T_{0i}^{\downarrow} + |\phi_{j}^{\downarrow}|^{2} T_{0i}^{\uparrow} \Big], \nonumber \\
\end{eqnarray}
where $z$ is the coordination number and for 2D systems $z=4$. We further define the superfluid order parameter vector $\Phi^{\dagger} = (\phi_{i}^{\uparrow \dagger}, \phi_{i}^{\downarrow \dagger}, \phi_{j}^{\uparrow \dagger}, \phi_{j}^{\downarrow \dagger})$, the perturbed ground-state energy is expressed in the matrix form as $E^{(2)} = \Phi^{\dagger} \textbf{M} \Phi$. Explicitly, 
\begin{equation}
\begin{aligned}
E^{(2)} &= \Phi^{\dagger} 
\begin{pmatrix}
\mathcal{A}_{j}^{\uparrow\downarrow} & 0 & zJ & \gamma \\
0 & \mathcal{A}_{j}^{\downarrow\uparrow} & -\gamma & zJ \\
zJ & -\gamma & \mathcal{A}_{i}^{\uparrow\downarrow} & 0 \\
\gamma & zJ & 0 & \mathcal{A}_{i}^{\downarrow\uparrow}
\end{pmatrix} \Phi \\
&= K_{1} \lvert \phi^{\uparrow}_{i} \rvert^{2}
+ K_{2} \lvert \phi^{\downarrow}_{i} \rvert^{2}
+ K_{3} \lvert \phi^{\uparrow}_{j} \rvert^{2}
+ K_{4} \lvert \phi^{\downarrow}_{j} \rvert^{2},
\label{ener_unper}
\end{aligned}
\end{equation}
where the diagonal elements are defined as
\begin{equation*}
\mathcal{A}_{k}^{\sigma\sigma'} = z^{2}J^{2}T_{0k}^{\sigma} + 2\gamma^{2}T_{0k}^{\sigma'}.
\end{equation*}
Here, $K_1, K_2, K_3$ and $K_4$ are the eigenvalues of matrix $\mathbf{M}$. The coefficients in the diagonal elements $T_{0i}^{\sigma}$ and $T_{0j}^{\sigma}$ represent the effective hopping contributions. For two sublattices $i$ and $j$, these are given by 
\begin{widetext}
\onecolumngrid
\begin{equation}
\begin{aligned}
T_{0i}^{\sigma} = \frac{1}{J_{0ci}^{\sigma}} = \;& \frac{n_i^{\sigma}+1}{U n_i^{\sigma} + U_{\uparrow\downarrow} n_i^{\sigma'} + 2 V n_j^{\sigma} + 2 V(\theta) n_j^{\sigma} + 2 V_{\uparrow \downarrow} n_{j}^{\sigma'} + 2 V_{\uparrow \downarrow}(\theta) n_{j}^{\sigma'} - \mu} \\
&- \frac{n_i^{\sigma}}{U (n_i^{\sigma}-1) + U_{\uparrow\downarrow} n_i^{\sigma'} + 2 V n_j^{\sigma} + 2 V(\theta) n_j^{\sigma} + 2 V_{\uparrow \downarrow} n_{j}^{\sigma'} + 2 V_{\uparrow \downarrow}(\theta) n_{j}^{\sigma'} - \mu},
\end{aligned}
\label{Tisite}
\end{equation}
\begin{equation}
\begin{aligned}
T_{0j}^{\sigma} = \frac{1}{J_{0cj}^{\sigma}} = \;& \frac{n_j^{\sigma}+1}{U n_j^{\sigma} + U_{\uparrow\downarrow} n_j^{\sigma'} + 2 V n_i^{\sigma} + 2 V(\theta) n_i^{\sigma} + 2 V_{\uparrow \downarrow} n_{i}^{\sigma'} + 2 V_{\uparrow \downarrow}(\theta) n_{i}^{\sigma'}  - \mu} \\
&- \frac{n_j^{\sigma}}{U (n_j^{\sigma}-1) + U_{\uparrow\downarrow} n_j^{\sigma'} + 2 V n_i^{\sigma} + 2 V(\theta) n_i^{\sigma} + 2 V_{\uparrow \downarrow} n_{i}^{\sigma'} + 2 V_{\uparrow \downarrow}(\theta) n_{i}^{\sigma'} - \mu}.
\end{aligned}
\label{Tjsite}
\end{equation}
\end{widetext}
\vspace{1.5em}
\twocolumngrid
Here, $J^{\uparrow}_{0ci}$ and $J^{\uparrow}_{0cj}$ denote the critical hopping amplitudes at sublattices $i$ and $j$, respectively, corresponding to the DW - SF(SS) quantum phase transition in the absence of spin-orbit coupling, i.e. $\gamma = 0$. In the presence of an intraspin nearest-neighbour interaction ($\zeta = 0$), the occupation numbers in the DW phase satisfy 
$n^{\uparrow}_{i} = n^{\downarrow}_{j}$ and $n^{\downarrow}_{i} = n^{\uparrow}_{j}$. As a result, the corresponding coefficients obey $T^{\uparrow}_{0i} = T^{\downarrow}_{0j}$ and 
$T^{\uparrow}_{0j} = T^{\downarrow}_{0i}$. Consequently, we obtain $K_{1} = K_{4}$ and $K_{2} = K_{3}$. These are given as 
\begin{subequations}
\begin{eqnarray}
  K_1 = K_4 &=& \frac{1}{2} \left[ (z^2 J^2 + 2\gamma^2)(T_{0i}^\uparrow + T_{0j}^\uparrow) + \sqrt{W} \right],~~~~~~~~~~\\
  K_2 = K_3 &=& \frac{1}{2} \left[ (z^2 J^2 + 2\gamma^2)(T_{0i}^\uparrow + T_{0j}^\uparrow) - \sqrt{W} \right],~~~~~~~~~~
\end{eqnarray}
\end{subequations}
where the effective hopping coefficients are defined as above. The discriminant $W$ is 
\begin{equation}
\begin{aligned}
W = &\Big[ \left( z^{2} J^{2} + 2 \gamma^{2} \right) \left( T_{0i}^{\uparrow} + T_{0j}^{\uparrow} \right) \Big]^{2} \nonumber \\
&- 4 \Big[ - \gamma^{2} + \gamma \left(z^{2} J^{2} - 2 \gamma^{2} \right) \left( T_{0i}^{\uparrow} - T_{0j}^{\uparrow} \right) \nonumber \\
&+ \left( 4 \gamma^{4} + z^{4} J^{4} \right) T_{0i}^{\uparrow} T_{0j}^{\uparrow} - z^{2} J^{2} \nonumber \\
&+ 2 \gamma^{2} z^{2} J^{2} \left( (T_{0i}^{\uparrow})^2 + (T_{0j}^{\uparrow})^2 \right) \Big].
\end{aligned}
\end{equation}
To determine the ground-state phase, we minimize the second-order energy correction $E^{(2)}_{(n^{\uparrow}_{i}, n^{\downarrow}_{i})}$ with respect to the SF order parameters of the components 
$\ket{\uparrow}$ and $\ket{\downarrow}$. This leads to the following conditions
\begin{equation}
\frac{\partial E^{(2)}}{\partial \phi^{\uparrow}_{i}} = 0,
\qquad
\frac{\partial E^{(2)}}{\partial \phi^{\downarrow}_{i}} = 0,
\end{equation}
which implies $K_{1} = K_{2} = 0$. The resulting condition therefore determines the CBDW-SF(SS) phase boundary. This is applicable when the dipoles are isotropic (no tilt) or tilted with respect to the polarization axis at angles below the magic angle. The critical hopping of the phase boundary is   
\begin{equation}
\begin{aligned}
z^{2} J^{2} + \gamma^{2} = \;& \left( z^{4} J^{4} + 4 \gamma^{4} \right) T_{0i}^{\uparrow} T_{0j}^{\uparrow} + 2 z^{2} J^{2} \gamma^{2} \Big[ (T_{0i}^{\uparrow})^{2} \\ 
&+ (T_{0j}^{\uparrow})^{2} \Big] 
 + \gamma \left( z^{2} J^{2} - 2 \gamma^{2} \right) \left( T_{0i}^{\uparrow} - T_{0j}^{\uparrow} \right),
\end{aligned}
\end{equation}
where the coefficients $T_{0i}^{\sigma}$ and $T_{0j}^{\sigma}$ are obtained from Eq.~(\ref{Tisite}) and Eq.(\ref{Tjsite}), respectively. 

\section{SDW-SF(SS) phase boundary}
\label{sec-II}
For stripe-ordered phase, the unperturbed ground-state energy at sublattices $i$ and $j$ remains the same as in Eq.~(\ref{ener_up}), and the perturbed Hamiltonian is redefined as in Eq.~(\ref{ham_pert_stripe}). The second-order perturbed ground-state energy for the striped-order phase transitions is 
\begin{equation}
\begin{aligned}
E^{(2)} =&\frac{z^{2}J^{2}}{4} \Big[ ( |\phi_j^{x \uparrow }|^{2} + |\phi_i^{y \uparrow}|^{2} ) T_{0i}^{\uparrow} + ( |\phi_i^{x \uparrow}|^{2} + |\phi_j^{y \uparrow}|^{2} ) T_{0j}^{\uparrow} \\
&+ ( |\phi_j^{x \downarrow}|^{2} + |\phi_i^{y\downarrow}|^{2} ) T_{0i}^{\downarrow} + ( |\phi_i^{x \downarrow}|^{2} + |\phi_j^{y \downarrow}|^{2} ) T_{0j}^{\downarrow} \Big] \\
&+ \frac{zJ}{2} \Big[2( \phi_i^{x \uparrow}\phi_j^{x \uparrow} + \phi_i^{x \downarrow}\phi_j^{x \downarrow} ) \\
&+ ( |\phi_i^{y \uparrow}|^{2} + |\phi_j^{y \uparrow}|^{2} + |\phi_i^{y \downarrow}|^{2} + |\phi_j^{y \downarrow}|^{2} )\Big] \\
&+ \gamma^{2} \Big[ ( |\phi_j^{x \downarrow}|^{2} + |\phi_i^{y \downarrow}|^{2} ) T_{0i}^{\uparrow} + ( |\phi_j^{x \uparrow }|^{2} + |\phi_i^{y \uparrow}|^{2} ) T_{0i}^{\downarrow} \\
&+ ( |\phi_i^{x \downarrow}|^{2} + |\phi_j^{y \downarrow }|^{2} ) T_{0j}^{\uparrow} + ( |\phi_i^{x \uparrow}|^{2} + |\phi_j^{y \uparrow}|^{2} ) T_{0j}^{\downarrow} \Big] \\
&+ 2 \gamma ( \phi_i^{x \uparrow}\phi_j^{x \downarrow} - \phi_j^{x \uparrow}\phi_i^{x \downarrow} ).
\end{aligned}
\end{equation}
As in the density-wave phase, the occupation number are $n_{i}^{\uparrow}$ = $n_{j}^{\downarrow}$ and $n_{i}^{\downarrow}$ = $n_{j}^{\uparrow}$, thus applying the conditions:
$T_{0i}^{\uparrow} = T_{0j}^{\downarrow}$ and $T_{0j}^{\uparrow} = T_{0i}^{\downarrow}$. We define 
\begin{equation}
  a = \frac{z^{2}J^{2}}{4} T_{0j}^{\uparrow} + \gamma^{2} T_{0i}^{\uparrow}, \quad
  b = \frac{z^{2}J^{2}}{4} T_{0i}^{\uparrow} + \gamma^{2} T_{0j}^{\uparrow}.   
\end{equation}
The perturbed energy is expressed in the matrix form as $E^{(2)} = \Phi^{\dagger} \mathbf{M} \Phi$.  In terms of $a$ and $b$ parameters, the matrix $\mathbf{M}$ is 
\begin{equation}
\hspace{-0.5cm}
\mathbf{M} =
\resizebox{\columnwidth}{!}{$
\begin{pmatrix}
a & 0 & 0 & 0 & zJ/2 & \gamma & 0 & 0 \\
0 & b & 0 & 0 & -\gamma & zJ/2 & 0 & 0 \\
0 & 0 & b+zJ/2 & 0 & 0 & 0 & 0 & 0 \\
0 & 0 & 0 & a+zJ/2 & 0 & 0 & 0 & 0 \\
zJ/2 & -\gamma & 0 & 0 & b & 0 & 0 & 0 \\
\gamma & zJ/2 & 0 & 0 & 0 & a & 0 & 0 \\
0 & 0 & 0 & 0 & 0 & 0 & a+zJ/2 & 0 \\
0 & 0 & 0 & 0 & 0 & 0 & 0 & b+zJ/2
\end{pmatrix}.
$}
\end{equation}
The correction in energy can be written as
\begin{equation*}
\begin{aligned}
E^{(2)} = \;& K_{1} \lvert \phi^{x \uparrow}_{i} \rvert^{2}
+ K_{2} \lvert \phi^{x \downarrow}_{i} \rvert^{2}
+ K_{3} \lvert \phi^{y \uparrow}_{i} \rvert^{2}
+ K_{4} \lvert \phi^{y \downarrow}_{i} \rvert^{2} \\
& + K_{5} \lvert \phi^{x \uparrow}_{j} \rvert^{2}
+ K_{6} \lvert \phi^{x \downarrow}_{j} \rvert^{2}
+ K_{7} \lvert \phi^{y \uparrow}_{j} \rvert^{2}
+ K_{8} \lvert \phi^{y \downarrow}_{j} \rvert^{2},
\end{aligned}
\end{equation*}
\noindent
where $\Phi^{\dagger} = \left( \phi_{i}^{x \uparrow \dagger}, \phi_{i}^{x \downarrow \dagger}, \phi_{i}^{y \uparrow \dagger}, \phi_{i}^{y \downarrow \dagger}, \phi_{j}^{x \uparrow \dagger}, \phi_{j}^{x \downarrow \dagger}, \phi_{j}^{y \uparrow \dagger}, \phi_{j}^{y \downarrow \dagger} \right)$ represents the basis vectors, and $K_1, K_2, \ldots, K_8$ are the eigenvalues of the matrix $\mathbf{M}$. These eigenvalues are
\begin{align*}
K_1 &= K_2 = a + zJ/2, \\
K_3 &= K_4 = b + zJ/2, \\
K_5 &= \frac{1}{2} \left[ a + b  - \sqrt{(a-b)^2 + 4(a\gamma - b\gamma + \gamma^2 + z^{2}J^{2}/4)} \right], \\
K_6 &= \frac{1}{2} \left[ a + b  + \sqrt{(a-b)^2 + 4(a\gamma - b\gamma + \gamma^2 + z^{2}J^{2}/4)} \right], \\
K_7 &= \frac{1}{2} \left[ a + b  - \sqrt{(a-b)^2 - 4(a\gamma - b\gamma - \gamma^2 - z^{2}J^{2}/4)} \right], \\
K_8 &= \frac{1}{2} \left[ a + b  + \sqrt{(a-b)^2 - 4(a\gamma - b\gamma - \gamma^2 - z^{2}J^{2}/4)} \right].
\end{align*}
To determine the ground-state phase, we minimize the second-order energy correction $E^{(2)}_{(n^{\uparrow}_{i}, n^{\downarrow}_{i})}$ with respect to the corresponding order parameters. The minimization condition requires $K_1, K_2, \ldots, K_8 = 0$. To obtain the phase-boundary condition, we restrict our analysis to $K_1$ and $K_3$, which are given by
\begin{subequations}
\begin{eqnarray}
K_1 &=& \frac{z^2J^2}{4} T_{0j}^{\uparrow} + \gamma^{2} T_{0i}^{\uparrow} + \frac{zJ}{2} = 0, \\
K_3 &=& \frac{z^2J^2}{4} T_{0i}^{\uparrow} + \gamma^{2} T_{0j}^{\uparrow} + \frac{zJ}{2} = 0.
\end{eqnarray}
\end{subequations}
By adding the above two (sub)equations, we obtain the critical hopping for the SDW-SF(SS) quantum phase transitions. The critical hopping read as 
\begin{equation}
\left(\frac{z^2J^2}{4} + \gamma^{2}\right)(T_{0i}^{\uparrow} + T_{0j}^{\uparrow}) = \frac{zJ}{2},
\end{equation}
where the effective hoppings are defined in Eq.\eqref{Tisite}. The above condition applies to dipolar bosons when the tilt of the dipoles is above the magic angle. 

\section{cDW-SF(SS) \& cMI-SF(SS) phase boundary}
\label{sec-III}
For correlated phases, the perturbed Hamiltonian remains the same as discussed previously Eq~(\ref{ham_per}). The correction in the ground-state energy is given by Eq.~(\ref{ener_unper}). 
In the presence of both intra- and interspin nearest-neighbour interactions, the cMI and cDW phases are characterized by equal spin occupations, $n_i^{\uparrow} = n_i^{\downarrow}$ and $n_j^{\uparrow} = n_j^{\downarrow}$. At the cMI-SF(SS) or cDW-SF(SS) phase boundary, the hopping amplitude satisfies $T_{0i}^{\uparrow} = T_{0i}^{\downarrow}$ and $T_{0j}^{\uparrow} = T_{0j}^{\downarrow}$, which leads to a pairwise degeneracy of the eigenvalues, $K_1 = K_2$ and $K_3 = K_4$. The corresponding eigenvalues are
\begin{subequations}
\begin{eqnarray}
K_1 = K_2 &=& \frac{1}{2} \left[ (z^2 J^2 + 2\gamma^2)(T_{0i}^\uparrow + T_{0j}^\uparrow) + \sqrt{W^{\prime}} \right],~~~~~~~~~~\\
K_3 = K_4 &=& \frac{1}{2} \left[ (z^2 J^2 + 2\gamma^2)(T_{0i}^\uparrow + T_{0j}^\uparrow) - \sqrt{W^{\prime}} \right],~~~~~~~~~~
\end{eqnarray}
\end{subequations}
where $W^{\prime}$ = $[(z^{2} J^{2} + 2 \gamma^{2})(T_{0i}^{\uparrow} + T_{0j}^{\uparrow})]^{2} - 4(z^{2} J^{2} + 2 \gamma^{2})^{2} T_{0i}^{\uparrow} T_{0j}^{\uparrow} + 4 \gamma^{2} + 4 z^{2} J^{2}$. Following the same procedure as discussed earlier, minimizing the second-order energy correction $E^{(2)}_{(n_{i}^{\uparrow}, n_{i}^{\downarrow})}$ with respect to the order parameter and get $K_{1} = K_{3} = 0$. The equation for the critical hopping of the cDW-SF(SS) and cMI-SF(SS) phase transition is obtained as
\begin{equation}
z^{2} J^{2} + \gamma^{2} = \left( z^{2} J^{2} + 2\gamma^{2} \right)^{2} T_{0i}^{\uparrow} T_{0j}^{\uparrow},
\end{equation}
where the effective hopping $T^{\sigma}_{0i}$ is defined in Eq.~(\ref{Tisite}).

\bibliography{dipsoc_tilt}{}

\begin{thebibliography}{73}%
\makeatletter
\providecommand \@ifxundefined [1]{%
 \@ifx{#1\undefined}
}%
\providecommand \@ifnum [1]{%
 \ifnum #1\expandafter \@firstoftwo
 \else \expandafter \@secondoftwo
 \fi
}%
\providecommand \@ifx [1]{%
 \ifx #1\expandafter \@firstoftwo
 \else \expandafter \@secondoftwo
 \fi
}%
\providecommand \natexlab [1]{#1}%
\providecommand \enquote  [1]{``#1''}%
\providecommand \bibnamefont  [1]{#1}%
\providecommand \bibfnamefont [1]{#1}%
\providecommand \citenamefont [1]{#1}%
\providecommand \href@noop [0]{\@secondoftwo}%
\providecommand \href [0]{\begingroup \@sanitize@url \@href}%
\providecommand \@href[1]{\@@startlink{#1}\@@href}%
\providecommand \@@href[1]{\endgroup#1\@@endlink}%
\providecommand \@sanitize@url [0]{\catcode `\\12\catcode `\$12\catcode
  `\&12\catcode `\#12\catcode `\^12\catcode `\_12\catcode `\%12\relax}%
\providecommand \@@startlink[1]{}%
\providecommand \@@endlink[0]{}%
\providecommand \url  [0]{\begingroup\@sanitize@url \@url }%
\providecommand \@url [1]{\endgroup\@href {#1}{\urlprefix }}%
\providecommand \urlprefix  [0]{URL }%
\providecommand \Eprint [0]{\href }%
\providecommand \doibase [0]{https://doi.org/}%
\providecommand \selectlanguage [0]{\@gobble}%
\providecommand \bibinfo  [0]{\@secondoftwo}%
\providecommand \bibfield  [0]{\@secondoftwo}%
\providecommand \translation [1]{[#1]}%
\providecommand \BibitemOpen [0]{}%
\providecommand \bibitemStop [0]{}%
\providecommand \bibitemNoStop [0]{.\EOS\space}%
\providecommand \EOS [0]{\spacefactor3000\relax}%
\providecommand \BibitemShut  [1]{\csname bibitem#1\endcsname}%
\let\auto@bib@innerbib\@empty
\bibitem [{\citenamefont {\ifmmode \check{Z}\else
  \v{Z}\fi{}uti\ifmmode~\acute{c}\else \'{c}\fi{}}\ \emph
  {et~al.}(2004)\citenamefont {\ifmmode \check{Z}\else
  \v{Z}\fi{}uti\ifmmode~\acute{c}\else \'{c}\fi{}}, \citenamefont {Fabian},\
  and\ \citenamefont {Das~Sarma}}]{zutic_04}%
  \BibitemOpen
  \bibfield  {author} {\bibinfo {author} {\bibfnamefont {I.}~\bibnamefont
  {\ifmmode \check{Z}\else \v{Z}\fi{}uti\ifmmode~\acute{c}\else \'{c}\fi{}}},
  \bibinfo {author} {\bibfnamefont {J.}~\bibnamefont {Fabian}},\ and\ \bibinfo
  {author} {\bibfnamefont {S.}~\bibnamefont {Das~Sarma}},\ }\href
  {https://doi.org/10.1103/RevModPhys.76.323} {\bibfield  {journal} {\bibinfo
  {journal} {Rev. Mod. Phys.}\ }\textbf {\bibinfo {volume} {76}},\ \bibinfo
  {pages} {323} (\bibinfo {year} {2004})}\BibitemShut {NoStop}%
\bibitem [{\citenamefont {AU~Lin}\ \emph {et~al.}(2011)\citenamefont {AU~Lin},
  \citenamefont {Jiménez-García},\ and\ \citenamefont {Spielman}}]{lin_11}%
  \BibitemOpen
  \bibfield  {author} {\bibinfo {author} {\bibfnamefont {Y.-J.}\ \bibnamefont
  {AU~Lin}}, \bibinfo {author} {\bibfnamefont {K.}~\bibnamefont
  {Jiménez-García}},\ and\ \bibinfo {author} {\bibfnamefont {I.~B.}\
  \bibnamefont {Spielman}},\ }\href {https://doi.org/10.1038/nature09887}
  {\bibfield  {journal} {\bibinfo  {journal} {Nature}\ }\textbf {\bibinfo
  {volume} {471}},\ \bibinfo {pages} {83} (\bibinfo {year} {2011})}\BibitemShut
  {NoStop}%
\bibitem [{\citenamefont {Galitski}\ and\ \citenamefont
  {Spielman}(2013)}]{galitski_13}%
  \BibitemOpen
  \bibfield  {author} {\bibinfo {author} {\bibfnamefont {V.}~\bibnamefont
  {Galitski}}\ and\ \bibinfo {author} {\bibfnamefont {I.~B.}\ \bibnamefont
  {Spielman}},\ }\href {https://doi.org/10.1038/nature11841} {\bibfield
  {journal} {\bibinfo  {journal} {Nature}\ }\textbf {\bibinfo {volume} {494}},\
  \bibinfo {pages} {49} (\bibinfo {year} {2013})}\BibitemShut {NoStop}%
\bibitem [{\citenamefont {Zhai}(2015)}]{zhai_15}%
  \BibitemOpen
  \bibfield  {author} {\bibinfo {author} {\bibfnamefont {H.}~\bibnamefont
  {Zhai}},\ }\href {https://doi.org/10.1088/0034-4885/78/2/026001} {\bibfield
  {journal} {\bibinfo  {journal} {Rep. Prog. Phys.}\ }\textbf {\bibinfo
  {volume} {78}},\ \bibinfo {pages} {026001} (\bibinfo {year}
  {2015})}\BibitemShut {NoStop}%
\bibitem [{\citenamefont {Hasan}\ and\ \citenamefont {Kane}(2010)}]{hasan_10}%
  \BibitemOpen
  \bibfield  {author} {\bibinfo {author} {\bibfnamefont {M.~Z.}\ \bibnamefont
  {Hasan}}\ and\ \bibinfo {author} {\bibfnamefont {C.~L.}\ \bibnamefont
  {Kane}},\ }\href {https://doi.org/10.1103/RevModPhys.82.3045} {\bibfield
  {journal} {\bibinfo  {journal} {Rev. Mod. Phys.}\ }\textbf {\bibinfo {volume}
  {82}},\ \bibinfo {pages} {3045} (\bibinfo {year} {2010})}\BibitemShut
  {NoStop}%
\bibitem [{\citenamefont {Qi}\ and\ \citenamefont {Zhang}(2011)}]{liang_11}%
  \BibitemOpen
  \bibfield  {author} {\bibinfo {author} {\bibfnamefont {X.-L.}\ \bibnamefont
  {Qi}}\ and\ \bibinfo {author} {\bibfnamefont {S.-C.}\ \bibnamefont {Zhang}},\
  }\href {https://doi.org/10.1103/RevModPhys.83.1057} {\bibfield  {journal}
  {\bibinfo  {journal} {Rev. Mod. Phys.}\ }\textbf {\bibinfo {volume} {83}},\
  \bibinfo {pages} {1057} (\bibinfo {year} {2011})}\BibitemShut {NoStop}%
\bibitem [{\citenamefont {Manchon}\ \emph {et~al.}(2015)\citenamefont
  {Manchon}, \citenamefont {Koo}, \citenamefont {Nitta}, \citenamefont
  {Frolov},\ and\ \citenamefont {Duine}}]{manchon_15}%
  \BibitemOpen
  \bibfield  {author} {\bibinfo {author} {\bibfnamefont {A.}~\bibnamefont
  {Manchon}}, \bibinfo {author} {\bibfnamefont {H.~C.}\ \bibnamefont {Koo}},
  \bibinfo {author} {\bibfnamefont {J.}~\bibnamefont {Nitta}}, \bibinfo
  {author} {\bibfnamefont {S.~M.}\ \bibnamefont {Frolov}},\ and\ \bibinfo
  {author} {\bibfnamefont {R.~A.}\ \bibnamefont {Duine}},\ }\href
  {https://doi.org/10.1038/nmat4360} {\bibfield  {journal} {\bibinfo  {journal}
  {Nat. Mater.}\ }\textbf {\bibinfo {volume} {14}},\ \bibinfo {pages} {871}
  (\bibinfo {year} {2015})}\BibitemShut {NoStop}%
\bibitem [{\citenamefont {Wu}\ \emph {et~al.}(2016)\citenamefont {Wu},
  \citenamefont {Zhang}, \citenamefont {Sun}, \citenamefont {Xu}, \citenamefont
  {Wang}, \citenamefont {Ji}, \citenamefont {Deng}, \citenamefont {Chen},
  \citenamefont {Liu},\ and\ \citenamefont {Pan}}]{wu_16}%
  \BibitemOpen
  \bibfield  {author} {\bibinfo {author} {\bibfnamefont {Z.}~\bibnamefont
  {Wu}}, \bibinfo {author} {\bibfnamefont {L.}~\bibnamefont {Zhang}}, \bibinfo
  {author} {\bibfnamefont {W.}~\bibnamefont {Sun}}, \bibinfo {author}
  {\bibfnamefont {X.-T.}\ \bibnamefont {Xu}}, \bibinfo {author} {\bibfnamefont
  {B.-Z.}\ \bibnamefont {Wang}}, \bibinfo {author} {\bibfnamefont {S.-C.}\
  \bibnamefont {Ji}}, \bibinfo {author} {\bibfnamefont {Y.}~\bibnamefont
  {Deng}}, \bibinfo {author} {\bibfnamefont {S.}~\bibnamefont {Chen}}, \bibinfo
  {author} {\bibfnamefont {X.-J.}\ \bibnamefont {Liu}},\ and\ \bibinfo {author}
  {\bibfnamefont {J.-W.}\ \bibnamefont {Pan}},\ }\href
  {https://doi.org/10.1126/science.aaf6689} {\bibfield  {journal} {\bibinfo
  {journal} {Science}\ }\textbf {\bibinfo {volume} {354}},\ \bibinfo {pages}
  {83} (\bibinfo {year} {2016})}\BibitemShut {NoStop}%
\bibitem [{\citenamefont {Huang}\ \emph {et~al.}(2016)\citenamefont {Huang},
  \citenamefont {Meng}, \citenamefont {Wang}, \citenamefont {Peng},
  \citenamefont {Zhang}, \citenamefont {Chen}, \citenamefont {Li},
  \citenamefont {Zhou},\ and\ \citenamefont {Zhang}}]{lianghui_16}%
  \BibitemOpen
  \bibfield  {author} {\bibinfo {author} {\bibfnamefont {L.}~\bibnamefont
  {Huang}}, \bibinfo {author} {\bibfnamefont {Z.}~\bibnamefont {Meng}},
  \bibinfo {author} {\bibfnamefont {P.}~\bibnamefont {Wang}}, \bibinfo {author}
  {\bibfnamefont {P.}~\bibnamefont {Peng}}, \bibinfo {author} {\bibfnamefont
  {S.-L.}\ \bibnamefont {Zhang}}, \bibinfo {author} {\bibfnamefont
  {L.}~\bibnamefont {Chen}}, \bibinfo {author} {\bibfnamefont {D.}~\bibnamefont
  {Li}}, \bibinfo {author} {\bibfnamefont {Q.}~\bibnamefont {Zhou}},\ and\
  \bibinfo {author} {\bibfnamefont {J.}~\bibnamefont {Zhang}},\ }\href
  {https://doi.org/https://doi.org/10.1038/nphys3672} {\bibfield  {journal}
  {\bibinfo  {journal} {Nat. Phys.}\ }\textbf {\bibinfo {volume} {12}},\
  \bibinfo {pages} {540} (\bibinfo {year} {2016})}\BibitemShut {NoStop}%
\bibitem [{\citenamefont {Sun}\ \emph {et~al.}(2018)\citenamefont {Sun},
  \citenamefont {Wang}, \citenamefont {Xu}, \citenamefont {Yi}, \citenamefont
  {Zhang}, \citenamefont {Wu}, \citenamefont {Deng}, \citenamefont {Liu},
  \citenamefont {Chen},\ and\ \citenamefont {Pan}}]{sun_18}%
  \BibitemOpen
  \bibfield  {author} {\bibinfo {author} {\bibfnamefont {W.}~\bibnamefont
  {Sun}}, \bibinfo {author} {\bibfnamefont {B.-Z.}\ \bibnamefont {Wang}},
  \bibinfo {author} {\bibfnamefont {X.-T.}\ \bibnamefont {Xu}}, \bibinfo
  {author} {\bibfnamefont {C.-R.}\ \bibnamefont {Yi}}, \bibinfo {author}
  {\bibfnamefont {L.}~\bibnamefont {Zhang}}, \bibinfo {author} {\bibfnamefont
  {Z.}~\bibnamefont {Wu}}, \bibinfo {author} {\bibfnamefont {Y.}~\bibnamefont
  {Deng}}, \bibinfo {author} {\bibfnamefont {X.-J.}\ \bibnamefont {Liu}},
  \bibinfo {author} {\bibfnamefont {S.}~\bibnamefont {Chen}},\ and\ \bibinfo
  {author} {\bibfnamefont {J.-W.}\ \bibnamefont {Pan}},\ }\href
  {https://doi.org/10.1103/PhysRevLett.121.150401} {\bibfield  {journal}
  {\bibinfo  {journal} {Phys. Rev. Lett.}\ }\textbf {\bibinfo {volume} {121}},\
  \bibinfo {pages} {150401} (\bibinfo {year} {2018})}\BibitemShut {NoStop}%
\bibitem [{\citenamefont {Li}\ \emph {et~al.}(2019)\citenamefont {Li},
  \citenamefont {Qu}, \citenamefont {Niffenegger}, \citenamefont {Wang},
  \citenamefont {He}, \citenamefont {Blasing}, \citenamefont {Olson},
  \citenamefont {Greene}, \citenamefont {Lyanda-Geller}, \citenamefont {Zhou},
  \citenamefont {Zhang},\ and\ \citenamefont {Chen}}]{hsun_19}%
  \BibitemOpen
  \bibfield  {author} {\bibinfo {author} {\bibfnamefont {C.-H.}\ \bibnamefont
  {Li}}, \bibinfo {author} {\bibfnamefont {C.}~\bibnamefont {Qu}}, \bibinfo
  {author} {\bibfnamefont {R.~J.}\ \bibnamefont {Niffenegger}}, \bibinfo
  {author} {\bibfnamefont {S.-J.}\ \bibnamefont {Wang}}, \bibinfo {author}
  {\bibfnamefont {M.}~\bibnamefont {He}}, \bibinfo {author} {\bibfnamefont
  {D.~B.}\ \bibnamefont {Blasing}}, \bibinfo {author} {\bibfnamefont {A.~J.}\
  \bibnamefont {Olson}}, \bibinfo {author} {\bibfnamefont {C.~H.}\ \bibnamefont
  {Greene}}, \bibinfo {author} {\bibfnamefont {Y.}~\bibnamefont
  {Lyanda-Geller}}, \bibinfo {author} {\bibfnamefont {Q.}~\bibnamefont {Zhou}},
  \bibinfo {author} {\bibfnamefont {C.}~\bibnamefont {Zhang}},\ and\ \bibinfo
  {author} {\bibfnamefont {Y.~P.}\ \bibnamefont {Chen}},\ }\href
  {https://doi.org/10.1038/s41467-018-08119-4} {\bibfield  {journal} {\bibinfo
  {journal} {Nat. Commun.}\ }\textbf {\bibinfo {volume} {10}},\ \bibinfo
  {pages} {375} (\bibinfo {year} {2019})}\BibitemShut {NoStop}%
\bibitem [{\citenamefont {Valdés-Curiel}\ \emph {et~al.}(2021)\citenamefont
  {Valdés-Curiel}, \citenamefont {Trypogeorgos}, \citenamefont {Liang},
  \citenamefont {Anderson},\ and\ \citenamefont {Spielman}}]{curiel_21}%
  \BibitemOpen
  \bibfield  {author} {\bibinfo {author} {\bibfnamefont {A.}~\bibnamefont
  {Valdés-Curiel}}, \bibinfo {author} {\bibfnamefont {D.}~\bibnamefont
  {Trypogeorgos}}, \bibinfo {author} {\bibfnamefont {Q.-Y.}\ \bibnamefont
  {Liang}}, \bibinfo {author} {\bibfnamefont {R.~P.}\ \bibnamefont
  {Anderson}},\ and\ \bibinfo {author} {\bibfnamefont {I.~B.}\ \bibnamefont
  {Spielman}},\ }\href {https://doi.org/10.1038/s41467-020-20762-4} {\bibfield
  {journal} {\bibinfo  {journal} {Nat. Commun.}\ }\textbf {\bibinfo {volume}
  {12}},\ \bibinfo {pages} {593} (\bibinfo {year} {2021})}\BibitemShut
  {NoStop}%
\bibitem [{\citenamefont {Wang}\ \emph {et~al.}(2021)\citenamefont {Wang},
  \citenamefont {Cheng}, \citenamefont {Wang}, \citenamefont {Zhang},
  \citenamefont {Lu}, \citenamefont {Yi}, \citenamefont {Niu}, \citenamefont
  {Deng}, \citenamefont {Liu}, \citenamefont {Chen},\ and\ \citenamefont
  {Pan}}]{wang_21}%
  \BibitemOpen
  \bibfield  {author} {\bibinfo {author} {\bibfnamefont {Z.-Y.}\ \bibnamefont
  {Wang}}, \bibinfo {author} {\bibfnamefont {X.-C.}\ \bibnamefont {Cheng}},
  \bibinfo {author} {\bibfnamefont {B.-Z.}\ \bibnamefont {Wang}}, \bibinfo
  {author} {\bibfnamefont {J.-Y.}\ \bibnamefont {Zhang}}, \bibinfo {author}
  {\bibfnamefont {Y.-H.}\ \bibnamefont {Lu}}, \bibinfo {author} {\bibfnamefont
  {C.-R.}\ \bibnamefont {Yi}}, \bibinfo {author} {\bibfnamefont
  {S.}~\bibnamefont {Niu}}, \bibinfo {author} {\bibfnamefont {Y.}~\bibnamefont
  {Deng}}, \bibinfo {author} {\bibfnamefont {X.-J.}\ \bibnamefont {Liu}},
  \bibinfo {author} {\bibfnamefont {S.}~\bibnamefont {Chen}},\ and\ \bibinfo
  {author} {\bibfnamefont {J.-W.}\ \bibnamefont {Pan}},\ }\href
  {https://doi.org/10.1126/science.abc0105} {\bibfield  {journal} {\bibinfo
  {journal} {Science}\ }\textbf {\bibinfo {volume} {372}},\ \bibinfo {pages}
  {271} (\bibinfo {year} {2021})}\BibitemShut {NoStop}%
\bibitem [{\citenamefont {Ren}\ \emph {et~al.}(2022)\citenamefont {Ren},
  \citenamefont {Liu}, \citenamefont {Zhao}, \citenamefont {He}, \citenamefont
  {Pak}, \citenamefont {Li},\ and\ \citenamefont {Jo}}]{zejian_22}%
  \BibitemOpen
  \bibfield  {author} {\bibinfo {author} {\bibfnamefont {Z.}~\bibnamefont
  {Ren}}, \bibinfo {author} {\bibfnamefont {D.}~\bibnamefont {Liu}}, \bibinfo
  {author} {\bibfnamefont {E.}~\bibnamefont {Zhao}}, \bibinfo {author}
  {\bibfnamefont {C.}~\bibnamefont {He}}, \bibinfo {author} {\bibfnamefont
  {K.~K.}\ \bibnamefont {Pak}}, \bibinfo {author} {\bibfnamefont
  {J.}~\bibnamefont {Li}},\ and\ \bibinfo {author} {\bibfnamefont {G.-B.}\
  \bibnamefont {Jo}},\ }\href {https://doi.org/10.1038/s41567-021-01491-x}
  {\bibfield  {journal} {\bibinfo  {journal} {Nat. Phys.}\ }\textbf {\bibinfo
  {volume} {18}},\ \bibinfo {pages} {385} (\bibinfo {year} {2022})}\BibitemShut
  {NoStop}%
\bibitem [{\citenamefont {Liang}\ \emph {et~al.}(2024)\citenamefont {Liang},
  \citenamefont {Dong}, \citenamefont {Pan}, \citenamefont {Wang},
  \citenamefont {Li}, \citenamefont {Yang}, \citenamefont {Yi},\ and\
  \citenamefont {Yan}}]{qian_24}%
  \BibitemOpen
  \bibfield  {author} {\bibinfo {author} {\bibfnamefont {Q.}~\bibnamefont
  {Liang}}, \bibinfo {author} {\bibfnamefont {Z.}~\bibnamefont {Dong}},
  \bibinfo {author} {\bibfnamefont {J.-S.}\ \bibnamefont {Pan}}, \bibinfo
  {author} {\bibfnamefont {H.}~\bibnamefont {Wang}}, \bibinfo {author}
  {\bibfnamefont {H.}~\bibnamefont {Li}}, \bibinfo {author} {\bibfnamefont
  {Z.}~\bibnamefont {Yang}}, \bibinfo {author} {\bibfnamefont {W.}~\bibnamefont
  {Yi}},\ and\ \bibinfo {author} {\bibfnamefont {B.}~\bibnamefont {Yan}},\
  }\href {https://doi.org/10.1038/s41567-024-02644-4} {\bibfield  {journal}
  {\bibinfo  {journal} {Nat. Phys.}\ }\textbf {\bibinfo {volume} {20}},\
  \bibinfo {pages} {1738} (\bibinfo {year} {2024})}\BibitemShut {NoStop}%
\bibitem [{\citenamefont {Lewenstein}\ \emph {et~al.}(2007)\citenamefont
  {Lewenstein}, \citenamefont {Sanpera}, \citenamefont {Ahufinger},
  \citenamefont {Damski}, \citenamefont {Sen(De)},\ and\ \citenamefont
  {Sen}}]{lewenstein_07}%
  \BibitemOpen
  \bibfield  {author} {\bibinfo {author} {\bibfnamefont {M.}~\bibnamefont
  {Lewenstein}}, \bibinfo {author} {\bibfnamefont {A.}~\bibnamefont {Sanpera}},
  \bibinfo {author} {\bibfnamefont {V.}~\bibnamefont {Ahufinger}}, \bibinfo
  {author} {\bibfnamefont {B.}~\bibnamefont {Damski}}, \bibinfo {author}
  {\bibfnamefont {A.}~\bibnamefont {Sen(De)}},\ and\ \bibinfo {author}
  {\bibfnamefont {U.}~\bibnamefont {Sen}},\ }\href
  {https://doi.org/10.1080/00018730701223200} {\bibfield  {journal} {\bibinfo
  {journal} {Adv. Phys.}\ }\textbf {\bibinfo {volume} {56}},\ \bibinfo {pages}
  {243} (\bibinfo {year} {2007})}\BibitemShut {NoStop}%
\bibitem [{\citenamefont {Boninsegni}\ and\ \citenamefont
  {Prokof'ev}(2012)}]{boninsegni_12}%
  \BibitemOpen
  \bibfield  {author} {\bibinfo {author} {\bibfnamefont {M.}~\bibnamefont
  {Boninsegni}}\ and\ \bibinfo {author} {\bibfnamefont {N.~V.}\ \bibnamefont
  {Prokof'ev}},\ }\href {https://doi.org/10.1103/RevModPhys.84.759} {\bibfield
  {journal} {\bibinfo  {journal} {Rev. Mod. Phys.}\ }\textbf {\bibinfo {volume}
  {84}},\ \bibinfo {pages} {759} (\bibinfo {year} {2012})}\BibitemShut
  {NoStop}%
\bibitem [{\citenamefont {Chomaz}\ \emph {et~al.}(2022)\citenamefont {Chomaz},
  \citenamefont {Ferrier-Barbut}, \citenamefont {Ferlaino}, \citenamefont
  {Laburthe-Tolra}, \citenamefont {Lev},\ and\ \citenamefont
  {Pfau}}]{chomaz_22}%
  \BibitemOpen
  \bibfield  {author} {\bibinfo {author} {\bibfnamefont {L.}~\bibnamefont
  {Chomaz}}, \bibinfo {author} {\bibfnamefont {I.}~\bibnamefont
  {Ferrier-Barbut}}, \bibinfo {author} {\bibfnamefont {F.}~\bibnamefont
  {Ferlaino}}, \bibinfo {author} {\bibfnamefont {B.}~\bibnamefont
  {Laburthe-Tolra}}, \bibinfo {author} {\bibfnamefont {B.~L.}\ \bibnamefont
  {Lev}},\ and\ \bibinfo {author} {\bibfnamefont {T.}~\bibnamefont {Pfau}},\
  }\href {https://doi.org/10.1088/1361-6633/aca814} {\bibfield  {journal}
  {\bibinfo  {journal} {Rep. Prog. Phys.}\ }\textbf {\bibinfo {volume} {86}},\
  \bibinfo {pages} {026401} (\bibinfo {year} {2022})}\BibitemShut {NoStop}%
\bibitem [{\citenamefont {Deng}\ \emph {et~al.}(2012)\citenamefont {Deng},
  \citenamefont {Cheng}, \citenamefont {Jing}, \citenamefont {Sun},\ and\
  \citenamefont {Yi}}]{deng_12}%
  \BibitemOpen
  \bibfield  {author} {\bibinfo {author} {\bibfnamefont {Y.}~\bibnamefont
  {Deng}}, \bibinfo {author} {\bibfnamefont {J.}~\bibnamefont {Cheng}},
  \bibinfo {author} {\bibfnamefont {H.}~\bibnamefont {Jing}}, \bibinfo {author}
  {\bibfnamefont {C.-P.}\ \bibnamefont {Sun}},\ and\ \bibinfo {author}
  {\bibfnamefont {S.}~\bibnamefont {Yi}},\ }\href
  {https://doi.org/10.1103/PhysRevLett.108.125301} {\bibfield  {journal}
  {\bibinfo  {journal} {Phys. Rev. Lett.}\ }\textbf {\bibinfo {volume} {108}},\
  \bibinfo {pages} {125301} (\bibinfo {year} {2012})}\BibitemShut {NoStop}%
\bibitem [{\citenamefont {Gopalakrishnan}\ \emph {et~al.}(2013)\citenamefont
  {Gopalakrishnan}, \citenamefont {Martin},\ and\ \citenamefont
  {Demler}}]{gopalkrishnan_13}%
  \BibitemOpen
  \bibfield  {author} {\bibinfo {author} {\bibfnamefont {S.}~\bibnamefont
  {Gopalakrishnan}}, \bibinfo {author} {\bibfnamefont {I.}~\bibnamefont
  {Martin}},\ and\ \bibinfo {author} {\bibfnamefont {E.~A.}\ \bibnamefont
  {Demler}},\ }\href {https://doi.org/10.1103/PhysRevLett.111.185304}
  {\bibfield  {journal} {\bibinfo  {journal} {Phys. Rev. Lett.}\ }\textbf
  {\bibinfo {volume} {111}},\ \bibinfo {pages} {185304} (\bibinfo {year}
  {2013})}\BibitemShut {NoStop}%
\bibitem [{\citenamefont {Wilson}\ \emph {et~al.}(2013)\citenamefont {Wilson},
  \citenamefont {Anderson},\ and\ \citenamefont {Clark}}]{wilson_13}%
  \BibitemOpen
  \bibfield  {author} {\bibinfo {author} {\bibfnamefont {R.~M.}\ \bibnamefont
  {Wilson}}, \bibinfo {author} {\bibfnamefont {B.~M.}\ \bibnamefont
  {Anderson}},\ and\ \bibinfo {author} {\bibfnamefont {C.~W.}\ \bibnamefont
  {Clark}},\ }\href {https://doi.org/10.1103/PhysRevLett.111.185303} {\bibfield
   {journal} {\bibinfo  {journal} {Phys. Rev. Lett.}\ }\textbf {\bibinfo
  {volume} {111}},\ \bibinfo {pages} {185303} (\bibinfo {year}
  {2013})}\BibitemShut {NoStop}%
\bibitem [{\citenamefont {Han}\ \emph {et~al.}(2018)\citenamefont {Han},
  \citenamefont {Zhang}, \citenamefont {Wang}, \citenamefont {Jiang},
  \citenamefont {Zhang},\ and\ \citenamefont {Zhang}}]{wei_18}%
  \BibitemOpen
  \bibfield  {author} {\bibinfo {author} {\bibfnamefont {W.}~\bibnamefont
  {Han}}, \bibinfo {author} {\bibfnamefont {X.-F.}\ \bibnamefont {Zhang}},
  \bibinfo {author} {\bibfnamefont {D.-S.}\ \bibnamefont {Wang}}, \bibinfo
  {author} {\bibfnamefont {H.-F.}\ \bibnamefont {Jiang}}, \bibinfo {author}
  {\bibfnamefont {W.}~\bibnamefont {Zhang}},\ and\ \bibinfo {author}
  {\bibfnamefont {S.-G.}\ \bibnamefont {Zhang}},\ }\href
  {https://doi.org/10.1103/PhysRevLett.121.030404} {\bibfield  {journal}
  {\bibinfo  {journal} {Phys. Rev. Lett.}\ }\textbf {\bibinfo {volume} {121}},\
  \bibinfo {pages} {030404} (\bibinfo {year} {2018})}\BibitemShut {NoStop}%
\bibitem [{\citenamefont {van Otterlo}\ \emph {et~al.}(1995)\citenamefont {van
  Otterlo}, \citenamefont {Wagenblast}, \citenamefont {Baltin}, \citenamefont
  {Bruder}, \citenamefont {Fazio},\ and\ \citenamefont {Sch\"on}}]{otterlo_95}%
  \BibitemOpen
  \bibfield  {author} {\bibinfo {author} {\bibfnamefont {A.}~\bibnamefont {van
  Otterlo}}, \bibinfo {author} {\bibfnamefont {K.-H.}\ \bibnamefont
  {Wagenblast}}, \bibinfo {author} {\bibfnamefont {R.}~\bibnamefont {Baltin}},
  \bibinfo {author} {\bibfnamefont {C.}~\bibnamefont {Bruder}}, \bibinfo
  {author} {\bibfnamefont {R.}~\bibnamefont {Fazio}},\ and\ \bibinfo {author}
  {\bibfnamefont {G.}~\bibnamefont {Sch\"on}},\ }\href
  {https://doi.org/10.1103/PhysRevB.52.16176} {\bibfield  {journal} {\bibinfo
  {journal} {Phys. Rev. B}\ }\textbf {\bibinfo {volume} {52}},\ \bibinfo
  {pages} {16176} (\bibinfo {year} {1995})}\BibitemShut {NoStop}%
\bibitem [{\citenamefont {Batrouni}\ and\ \citenamefont
  {Scalettar}(2000)}]{batrouni_00}%
  \BibitemOpen
  \bibfield  {author} {\bibinfo {author} {\bibfnamefont {G.~G.}\ \bibnamefont
  {Batrouni}}\ and\ \bibinfo {author} {\bibfnamefont {R.~T.}\ \bibnamefont
  {Scalettar}},\ }\href {https://doi.org/10.1103/PhysRevLett.84.1599}
  {\bibfield  {journal} {\bibinfo  {journal} {Phys. Rev. Lett.}\ }\textbf
  {\bibinfo {volume} {84}},\ \bibinfo {pages} {1599} (\bibinfo {year}
  {2000})}\BibitemShut {NoStop}%
\bibitem [{\citenamefont {Sengupta}\ \emph {et~al.}(2005)\citenamefont
  {Sengupta}, \citenamefont {Pryadko}, \citenamefont {Alet}, \citenamefont
  {Troyer},\ and\ \citenamefont {Schmid}}]{sengupta_05}%
  \BibitemOpen
  \bibfield  {author} {\bibinfo {author} {\bibfnamefont {P.}~\bibnamefont
  {Sengupta}}, \bibinfo {author} {\bibfnamefont {L.~P.}\ \bibnamefont
  {Pryadko}}, \bibinfo {author} {\bibfnamefont {F.}~\bibnamefont {Alet}},
  \bibinfo {author} {\bibfnamefont {M.}~\bibnamefont {Troyer}},\ and\ \bibinfo
  {author} {\bibfnamefont {G.}~\bibnamefont {Schmid}},\ }\href
  {https://doi.org/10.1103/PhysRevLett.94.207202} {\bibfield  {journal}
  {\bibinfo  {journal} {Phys. Rev. Lett.}\ }\textbf {\bibinfo {volume} {94}},\
  \bibinfo {pages} {207202} (\bibinfo {year} {2005})}\BibitemShut {NoStop}%
\bibitem [{\citenamefont {Capogrosso-Sansone}\ \emph
  {et~al.}(2010)\citenamefont {Capogrosso-Sansone}, \citenamefont {Trefzger},
  \citenamefont {Lewenstein}, \citenamefont {Zoller},\ and\ \citenamefont
  {Pupillo}}]{sansone_10}%
  \BibitemOpen
  \bibfield  {author} {\bibinfo {author} {\bibfnamefont {B.}~\bibnamefont
  {Capogrosso-Sansone}}, \bibinfo {author} {\bibfnamefont {C.}~\bibnamefont
  {Trefzger}}, \bibinfo {author} {\bibfnamefont {M.}~\bibnamefont
  {Lewenstein}}, \bibinfo {author} {\bibfnamefont {P.}~\bibnamefont {Zoller}},\
  and\ \bibinfo {author} {\bibfnamefont {G.}~\bibnamefont {Pupillo}},\ }\href
  {https://doi.org/10.1103/PhysRevLett.104.125301} {\bibfield  {journal}
  {\bibinfo  {journal} {Phys. Rev. Lett.}\ }\textbf {\bibinfo {volume} {104}},\
  \bibinfo {pages} {125301} (\bibinfo {year} {2010})}\BibitemShut {NoStop}%
\bibitem [{\citenamefont {Ohgoe}\ \emph {et~al.}(2012)\citenamefont {Ohgoe},
  \citenamefont {Suzuki},\ and\ \citenamefont {Kawashima}}]{ohgoe_12}%
  \BibitemOpen
  \bibfield  {author} {\bibinfo {author} {\bibfnamefont {T.}~\bibnamefont
  {Ohgoe}}, \bibinfo {author} {\bibfnamefont {T.}~\bibnamefont {Suzuki}},\ and\
  \bibinfo {author} {\bibfnamefont {N.}~\bibnamefont {Kawashima}},\ }\href
  {https://doi.org/10.1103/PhysRevB.86.054520} {\bibfield  {journal} {\bibinfo
  {journal} {Phys. Rev. B}\ }\textbf {\bibinfo {volume} {86}},\ \bibinfo
  {pages} {054520} (\bibinfo {year} {2012})}\BibitemShut {NoStop}%
\bibitem [{\citenamefont {Zhang}\ \emph {et~al.}(2015)\citenamefont {Zhang},
  \citenamefont {Safavi-Naini}, \citenamefont {Rey},\ and\ \citenamefont
  {Capogrosso-Sansone}}]{zhang_15}%
  \BibitemOpen
  \bibfield  {author} {\bibinfo {author} {\bibfnamefont {C.}~\bibnamefont
  {Zhang}}, \bibinfo {author} {\bibfnamefont {A.}~\bibnamefont {Safavi-Naini}},
  \bibinfo {author} {\bibfnamefont {A.~M.}\ \bibnamefont {Rey}},\ and\ \bibinfo
  {author} {\bibfnamefont {B.}~\bibnamefont {Capogrosso-Sansone}},\ }\href
  {https://doi.org/10.1088/1367-2630/17/12/123014} {\bibfield  {journal}
  {\bibinfo  {journal} {New J. Phys.}\ }\textbf {\bibinfo {volume} {17}},\
  \bibinfo {pages} {123014} (\bibinfo {year} {2015})}\BibitemShut {NoStop}%
\bibitem [{\citenamefont {Bandyopadhyay}\ \emph {et~al.}(2019)\citenamefont
  {Bandyopadhyay}, \citenamefont {Bai}, \citenamefont {Pal}, \citenamefont
  {Suthar}, \citenamefont {Nath},\ and\ \citenamefont
  {Angom}}]{bandyopadhyay_19}%
  \BibitemOpen
  \bibfield  {author} {\bibinfo {author} {\bibfnamefont {S.}~\bibnamefont
  {Bandyopadhyay}}, \bibinfo {author} {\bibfnamefont {R.}~\bibnamefont {Bai}},
  \bibinfo {author} {\bibfnamefont {S.}~\bibnamefont {Pal}}, \bibinfo {author}
  {\bibfnamefont {K.}~\bibnamefont {Suthar}}, \bibinfo {author} {\bibfnamefont
  {R.}~\bibnamefont {Nath}},\ and\ \bibinfo {author} {\bibfnamefont
  {D.}~\bibnamefont {Angom}},\ }\href
  {https://doi.org/10.1103/PhysRevA.100.053623} {\bibfield  {journal} {\bibinfo
   {journal} {Phys. Rev. A}\ }\textbf {\bibinfo {volume} {100}},\ \bibinfo
  {pages} {053623} (\bibinfo {year} {2019})}\BibitemShut {NoStop}%
\bibitem [{\citenamefont {Wu}\ and\ \citenamefont {Tu}(2020)}]{wu_20}%
  \BibitemOpen
  \bibfield  {author} {\bibinfo {author} {\bibfnamefont {H.-K.}\ \bibnamefont
  {Wu}}\ and\ \bibinfo {author} {\bibfnamefont {W.-L.}\ \bibnamefont {Tu}},\
  }\href {https://doi.org/10.1103/PhysRevA.102.053306} {\bibfield  {journal}
  {\bibinfo  {journal} {Phys. Rev. A}\ }\textbf {\bibinfo {volume} {102}},\
  \bibinfo {pages} {053306} (\bibinfo {year} {2020})}\BibitemShut {NoStop}%
\bibitem [{\citenamefont {Zhang}\ \emph {et~al.}(2021)\citenamefont {Zhang},
  \citenamefont {Zhang}, \citenamefont {Yang},\ and\ \citenamefont
  {Capogrosso-Sansone}}]{zhang_21}%
  \BibitemOpen
  \bibfield  {author} {\bibinfo {author} {\bibfnamefont {C.}~\bibnamefont
  {Zhang}}, \bibinfo {author} {\bibfnamefont {J.}~\bibnamefont {Zhang}},
  \bibinfo {author} {\bibfnamefont {J.}~\bibnamefont {Yang}},\ and\ \bibinfo
  {author} {\bibfnamefont {B.}~\bibnamefont {Capogrosso-Sansone}},\ }\href
  {https://doi.org/10.1103/PhysRevA.103.043333} {\bibfield  {journal} {\bibinfo
   {journal} {Phys. Rev. A}\ }\textbf {\bibinfo {volume} {103}},\ \bibinfo
  {pages} {043333} (\bibinfo {year} {2021})}\BibitemShut {NoStop}%
\bibitem [{\citenamefont {Zhang}\ \emph {et~al.}(2022)\citenamefont {Zhang},
  \citenamefont {Zhang}, \citenamefont {Yang},\ and\ \citenamefont
  {Capogrosso-Sansone}}]{zhang_22}%
  \BibitemOpen
  \bibfield  {author} {\bibinfo {author} {\bibfnamefont {J.}~\bibnamefont
  {Zhang}}, \bibinfo {author} {\bibfnamefont {C.}~\bibnamefont {Zhang}},
  \bibinfo {author} {\bibfnamefont {J.}~\bibnamefont {Yang}},\ and\ \bibinfo
  {author} {\bibfnamefont {B.}~\bibnamefont {Capogrosso-Sansone}},\ }\href
  {https://doi.org/10.1103/PhysRevA.105.063302} {\bibfield  {journal} {\bibinfo
   {journal} {Phys. Rev. A}\ }\textbf {\bibinfo {volume} {105}},\ \bibinfo
  {pages} {063302} (\bibinfo {year} {2022})}\BibitemShut {NoStop}%
\bibitem [{\citenamefont {Nguyen}\ and\ \citenamefont
  {Boninsegni}(2022)}]{nguyen_22}%
  \BibitemOpen
  \bibfield  {author} {\bibinfo {author} {\bibfnamefont {P.~H.}\ \bibnamefont
  {Nguyen}}\ and\ \bibinfo {author} {\bibfnamefont {M.}~\bibnamefont
  {Boninsegni}},\ }\href {https://doi.org/10.1007/s10909-022-02793-x}
  {\bibfield  {journal} {\bibinfo  {journal} {J. Low Temp. Phys.}\ }\textbf
  {\bibinfo {volume} {209}},\ \bibinfo {pages} {34} (\bibinfo {year}
  {2022})}\BibitemShut {NoStop}%
\bibitem [{\citenamefont {Léonard}\ \emph {et~al.}(2017)\citenamefont
  {Léonard}, \citenamefont {Morales}, \citenamefont {Zupancic}, \citenamefont
  {Esslinger},\ and\ \citenamefont {Donner}}]{leonard_17}%
  \BibitemOpen
  \bibfield  {author} {\bibinfo {author} {\bibfnamefont {J.}~\bibnamefont
  {Léonard}}, \bibinfo {author} {\bibfnamefont {A.}~\bibnamefont {Morales}},
  \bibinfo {author} {\bibfnamefont {P.}~\bibnamefont {Zupancic}}, \bibinfo
  {author} {\bibfnamefont {T.}~\bibnamefont {Esslinger}},\ and\ \bibinfo
  {author} {\bibfnamefont {T.}~\bibnamefont {Donner}},\ }\href
  {https://doi.org/10.1038/nature21067} {\bibfield  {journal} {\bibinfo
  {journal} {Nature}\ }\textbf {\bibinfo {volume} {543}},\ \bibinfo {pages}
  {87} (\bibinfo {year} {2017})}\BibitemShut {NoStop}%
\bibitem [{\citenamefont {Li}\ \emph {et~al.}(2017)\citenamefont {Li},
  \citenamefont {Lee}, \citenamefont {Huang}, \citenamefont {Burchesky},
  \citenamefont {Shteynas}, \citenamefont {Top}, \citenamefont {Jamison},\ and\
  \citenamefont {Ketterle}}]{li_17}%
  \BibitemOpen
  \bibfield  {author} {\bibinfo {author} {\bibfnamefont {J.-R.}\ \bibnamefont
  {Li}}, \bibinfo {author} {\bibfnamefont {J.}~\bibnamefont {Lee}}, \bibinfo
  {author} {\bibfnamefont {W.}~\bibnamefont {Huang}}, \bibinfo {author}
  {\bibfnamefont {S.}~\bibnamefont {Burchesky}}, \bibinfo {author}
  {\bibfnamefont {B.}~\bibnamefont {Shteynas}}, \bibinfo {author}
  {\bibfnamefont {F.~C.}\ \bibnamefont {Top}}, \bibinfo {author} {\bibfnamefont
  {A.~O.}\ \bibnamefont {Jamison}},\ and\ \bibinfo {author} {\bibfnamefont
  {W.}~\bibnamefont {Ketterle}},\ }\href {https://doi.org/10.1038/nature21431}
  {\bibfield  {journal} {\bibinfo  {journal} {Nature}\ }\textbf {\bibinfo
  {volume} {543}},\ \bibinfo {pages} {91} (\bibinfo {year} {2017})}\BibitemShut
  {NoStop}%
\bibitem [{\citenamefont {Guo}\ \emph {et~al.}(2019)\citenamefont {Guo},
  \citenamefont {Böttcher}, \citenamefont {Hertkorn}, \citenamefont {Schmidt},
  \citenamefont {Wenzel}, \citenamefont {Büchler}, \citenamefont {Langen},\
  and\ \citenamefont {Pfau}}]{guo_19}%
  \BibitemOpen
  \bibfield  {author} {\bibinfo {author} {\bibfnamefont {M.}~\bibnamefont
  {Guo}}, \bibinfo {author} {\bibfnamefont {F.}~\bibnamefont {Böttcher}},
  \bibinfo {author} {\bibfnamefont {J.}~\bibnamefont {Hertkorn}}, \bibinfo
  {author} {\bibfnamefont {J.-N.}\ \bibnamefont {Schmidt}}, \bibinfo {author}
  {\bibfnamefont {M.}~\bibnamefont {Wenzel}}, \bibinfo {author} {\bibfnamefont
  {H.~P.}\ \bibnamefont {Büchler}}, \bibinfo {author} {\bibfnamefont
  {T.}~\bibnamefont {Langen}},\ and\ \bibinfo {author} {\bibfnamefont
  {T.}~\bibnamefont {Pfau}},\ }\href
  {https://doi.org/10.1038/s41586-019-1569-5} {\bibfield  {journal} {\bibinfo
  {journal} {Nature}\ }\textbf {\bibinfo {volume} {574}},\ \bibinfo {pages}
  {386} (\bibinfo {year} {2019})}\BibitemShut {NoStop}%
\bibitem [{\citenamefont {Tanzi}\ \emph {et~al.}(2019)\citenamefont {Tanzi},
  \citenamefont {Roccuzzo}, \citenamefont {Lucioni}, \citenamefont {Famà},
  \citenamefont {Fioretti}, \citenamefont {Gabbanini}, \citenamefont {Modugno},
  \citenamefont {Recati},\ and\ \citenamefont {Stringari}}]{tanzi_19}%
  \BibitemOpen
  \bibfield  {author} {\bibinfo {author} {\bibfnamefont {L.}~\bibnamefont
  {Tanzi}}, \bibinfo {author} {\bibfnamefont {S.~M.}\ \bibnamefont {Roccuzzo}},
  \bibinfo {author} {\bibfnamefont {E.}~\bibnamefont {Lucioni}}, \bibinfo
  {author} {\bibfnamefont {F.}~\bibnamefont {Famà}}, \bibinfo {author}
  {\bibfnamefont {A.}~\bibnamefont {Fioretti}}, \bibinfo {author}
  {\bibfnamefont {C.}~\bibnamefont {Gabbanini}}, \bibinfo {author}
  {\bibfnamefont {G.}~\bibnamefont {Modugno}}, \bibinfo {author} {\bibfnamefont
  {A.}~\bibnamefont {Recati}},\ and\ \bibinfo {author} {\bibfnamefont
  {S.}~\bibnamefont {Stringari}},\ }\href
  {https://doi.org/10.1038/s41586-019-1568-6} {\bibfield  {journal} {\bibinfo
  {journal} {Nature}\ }\textbf {\bibinfo {volume} {574}},\ \bibinfo {pages}
  {382} (\bibinfo {year} {2019})}\BibitemShut {NoStop}%
\bibitem [{\citenamefont {Natale}\ \emph {et~al.}(2019)\citenamefont {Natale},
  \citenamefont {van Bijnen}, \citenamefont {Patscheider}, \citenamefont
  {Petter}, \citenamefont {Mark}, \citenamefont {Chomaz},\ and\ \citenamefont
  {Ferlaino}}]{natale_19}%
  \BibitemOpen
  \bibfield  {author} {\bibinfo {author} {\bibfnamefont {G.}~\bibnamefont
  {Natale}}, \bibinfo {author} {\bibfnamefont {R.~M.~W.}\ \bibnamefont {van
  Bijnen}}, \bibinfo {author} {\bibfnamefont {A.}~\bibnamefont {Patscheider}},
  \bibinfo {author} {\bibfnamefont {D.}~\bibnamefont {Petter}}, \bibinfo
  {author} {\bibfnamefont {M.~J.}\ \bibnamefont {Mark}}, \bibinfo {author}
  {\bibfnamefont {L.}~\bibnamefont {Chomaz}},\ and\ \bibinfo {author}
  {\bibfnamefont {F.}~\bibnamefont {Ferlaino}},\ }\href
  {https://doi.org/10.1103/PhysRevLett.123.050402} {\bibfield  {journal}
  {\bibinfo  {journal} {Phys. Rev. Lett.}\ }\textbf {\bibinfo {volume} {123}},\
  \bibinfo {pages} {050402} (\bibinfo {year} {2019})}\BibitemShut {NoStop}%
\bibitem [{\citenamefont {Norcia}\ \emph {et~al.}(2021)\citenamefont {Norcia},
  \citenamefont {Politi}, \citenamefont {Klaus}, \citenamefont {Poli},
  \citenamefont {Sohmen}, \citenamefont {Mark}, \citenamefont {Bisset},
  \citenamefont {Santos},\ and\ \citenamefont {Ferlaino}}]{norcia_21}%
  \BibitemOpen
  \bibfield  {author} {\bibinfo {author} {\bibfnamefont {M.~A.}\ \bibnamefont
  {Norcia}}, \bibinfo {author} {\bibfnamefont {C.}~\bibnamefont {Politi}},
  \bibinfo {author} {\bibfnamefont {L.}~\bibnamefont {Klaus}}, \bibinfo
  {author} {\bibfnamefont {E.}~\bibnamefont {Poli}}, \bibinfo {author}
  {\bibfnamefont {M.}~\bibnamefont {Sohmen}}, \bibinfo {author} {\bibfnamefont
  {M.~J.}\ \bibnamefont {Mark}}, \bibinfo {author} {\bibfnamefont {R.~N.}\
  \bibnamefont {Bisset}}, \bibinfo {author} {\bibfnamefont {L.}~\bibnamefont
  {Santos}},\ and\ \bibinfo {author} {\bibfnamefont {F.}~\bibnamefont
  {Ferlaino}},\ }\href {https://doi.org/10.1038/s41586-021-03725-7} {\bibfield
  {journal} {\bibinfo  {journal} {Nature}\ }\textbf {\bibinfo {volume} {596}},\
  \bibinfo {pages} {357} (\bibinfo {year} {2021})}\BibitemShut {NoStop}%
\bibitem [{\citenamefont {Bland}\ \emph {et~al.}(2022)\citenamefont {Bland},
  \citenamefont {Poli}, \citenamefont {Politi}, \citenamefont {Klaus},
  \citenamefont {Norcia}, \citenamefont {Ferlaino}, \citenamefont {Santos},\
  and\ \citenamefont {Bisset}}]{bland_22}%
  \BibitemOpen
  \bibfield  {author} {\bibinfo {author} {\bibfnamefont {T.}~\bibnamefont
  {Bland}}, \bibinfo {author} {\bibfnamefont {E.}~\bibnamefont {Poli}},
  \bibinfo {author} {\bibfnamefont {C.}~\bibnamefont {Politi}}, \bibinfo
  {author} {\bibfnamefont {L.}~\bibnamefont {Klaus}}, \bibinfo {author}
  {\bibfnamefont {M.~A.}\ \bibnamefont {Norcia}}, \bibinfo {author}
  {\bibfnamefont {F.}~\bibnamefont {Ferlaino}}, \bibinfo {author}
  {\bibfnamefont {L.}~\bibnamefont {Santos}},\ and\ \bibinfo {author}
  {\bibfnamefont {R.~N.}\ \bibnamefont {Bisset}},\ }\href
  {https://doi.org/10.1103/PhysRevLett.128.195302} {\bibfield  {journal}
  {\bibinfo  {journal} {Phys. Rev. Lett.}\ }\textbf {\bibinfo {volume} {128}},\
  \bibinfo {pages} {195302} (\bibinfo {year} {2022})}\BibitemShut {NoStop}%
\bibitem [{\citenamefont {Kirkby}\ \emph {et~al.}(2024)\citenamefont {Kirkby},
  \citenamefont {Lee}, \citenamefont {Baillie}, \citenamefont {Bland},
  \citenamefont {Ferlaino}, \citenamefont {Blakie},\ and\ \citenamefont
  {Bisset}}]{kirkby_24}%
  \BibitemOpen
  \bibfield  {author} {\bibinfo {author} {\bibfnamefont {W.}~\bibnamefont
  {Kirkby}}, \bibinfo {author} {\bibfnamefont {A.-C.}\ \bibnamefont {Lee}},
  \bibinfo {author} {\bibfnamefont {D.}~\bibnamefont {Baillie}}, \bibinfo
  {author} {\bibfnamefont {T.}~\bibnamefont {Bland}}, \bibinfo {author}
  {\bibfnamefont {F.}~\bibnamefont {Ferlaino}}, \bibinfo {author}
  {\bibfnamefont {P.~B.}\ \bibnamefont {Blakie}},\ and\ \bibinfo {author}
  {\bibfnamefont {R.~N.}\ \bibnamefont {Bisset}},\ }\href
  {https://doi.org/10.1103/PhysRevLett.133.103401} {\bibfield  {journal}
  {\bibinfo  {journal} {Phys. Rev. Lett.}\ }\textbf {\bibinfo {volume} {133}},\
  \bibinfo {pages} {103401} (\bibinfo {year} {2024})}\BibitemShut {NoStop}%
\bibitem [{\citenamefont {Lagoin}\ \emph {et~al.}(2022)\citenamefont {Lagoin},
  \citenamefont {Bhattacharya}, \citenamefont {Grass}, \citenamefont
  {Chhajlany}, \citenamefont {Salamon}, \citenamefont {Baldwin}, \citenamefont
  {Pfeiffer}, \citenamefont {Lewenstein}, \citenamefont {Holzmann},\ and\
  \citenamefont {Dubin}}]{lagoin_22}%
  \BibitemOpen
  \bibfield  {author} {\bibinfo {author} {\bibfnamefont {C.}~\bibnamefont
  {Lagoin}}, \bibinfo {author} {\bibfnamefont {U.}~\bibnamefont
  {Bhattacharya}}, \bibinfo {author} {\bibfnamefont {T.}~\bibnamefont {Grass}},
  \bibinfo {author} {\bibfnamefont {R.~W.}\ \bibnamefont {Chhajlany}}, \bibinfo
  {author} {\bibfnamefont {T.}~\bibnamefont {Salamon}}, \bibinfo {author}
  {\bibfnamefont {K.}~\bibnamefont {Baldwin}}, \bibinfo {author} {\bibfnamefont
  {L.}~\bibnamefont {Pfeiffer}}, \bibinfo {author} {\bibfnamefont
  {M.}~\bibnamefont {Lewenstein}}, \bibinfo {author} {\bibfnamefont
  {M.}~\bibnamefont {Holzmann}},\ and\ \bibinfo {author} {\bibfnamefont
  {F.}~\bibnamefont {Dubin}},\ }\href
  {https://doi.org/10.1038/s41586-022-05123-z} {\bibfield  {journal} {\bibinfo
  {journal} {Nature}\ }\textbf {\bibinfo {volume} {609}},\ \bibinfo {pages}
  {485} (\bibinfo {year} {2022})}\BibitemShut {NoStop}%
\bibitem [{\citenamefont {Baier}\ \emph {et~al.}(2016)\citenamefont {Baier},
  \citenamefont {Mark}, \citenamefont {Petter}, \citenamefont {Aikawa},
  \citenamefont {Chomaz}, \citenamefont {Cai}, \citenamefont {Baranov},
  \citenamefont {Zoller},\ and\ \citenamefont {Ferlaino}}]{baier_16}%
  \BibitemOpen
  \bibfield  {author} {\bibinfo {author} {\bibfnamefont {S.}~\bibnamefont
  {Baier}}, \bibinfo {author} {\bibfnamefont {M.~J.}\ \bibnamefont {Mark}},
  \bibinfo {author} {\bibfnamefont {D.}~\bibnamefont {Petter}}, \bibinfo
  {author} {\bibfnamefont {K.}~\bibnamefont {Aikawa}}, \bibinfo {author}
  {\bibfnamefont {L.}~\bibnamefont {Chomaz}}, \bibinfo {author} {\bibfnamefont
  {Z.}~\bibnamefont {Cai}}, \bibinfo {author} {\bibfnamefont {M.}~\bibnamefont
  {Baranov}}, \bibinfo {author} {\bibfnamefont {P.}~\bibnamefont {Zoller}},\
  and\ \bibinfo {author} {\bibfnamefont {F.}~\bibnamefont {Ferlaino}},\ }\href
  {https://doi.org/10.1126/science.aac9812} {\bibfield  {journal} {\bibinfo
  {journal} {Science}\ }\textbf {\bibinfo {volume} {352}},\ \bibinfo {pages}
  {201} (\bibinfo {year} {2016})}\BibitemShut {NoStop}%
\bibitem [{\citenamefont {Wilson}\ \emph {et~al.}(2016)\citenamefont {Wilson},
  \citenamefont {Shirley},\ and\ \citenamefont {Natu}}]{wilson_16}%
  \BibitemOpen
  \bibfield  {author} {\bibinfo {author} {\bibfnamefont {R.~M.}\ \bibnamefont
  {Wilson}}, \bibinfo {author} {\bibfnamefont {W.~E.}\ \bibnamefont
  {Shirley}},\ and\ \bibinfo {author} {\bibfnamefont {S.~S.}\ \bibnamefont
  {Natu}},\ }\href {https://doi.org/10.1103/PhysRevA.93.011605} {\bibfield
  {journal} {\bibinfo  {journal} {Phys. Rev. A}\ }\textbf {\bibinfo {volume}
  {93}},\ \bibinfo {pages} {011605} (\bibinfo {year} {2016})}\BibitemShut
  {NoStop}%
\bibitem [{\citenamefont {Guan}\ \emph {et~al.}(2019)\citenamefont {Guan},
  \citenamefont {Fan}, \citenamefont {Zhou}, \citenamefont {Chen},\ and\
  \citenamefont {Jia}}]{guan_19}%
  \BibitemOpen
  \bibfield  {author} {\bibinfo {author} {\bibfnamefont {X.}~\bibnamefont
  {Guan}}, \bibinfo {author} {\bibfnamefont {J.}~\bibnamefont {Fan}}, \bibinfo
  {author} {\bibfnamefont {X.}~\bibnamefont {Zhou}}, \bibinfo {author}
  {\bibfnamefont {G.}~\bibnamefont {Chen}},\ and\ \bibinfo {author}
  {\bibfnamefont {S.}~\bibnamefont {Jia}},\ }\href
  {https://doi.org/10.1103/PhysRevA.100.013617} {\bibfield  {journal} {\bibinfo
   {journal} {Phys. Rev. A}\ }\textbf {\bibinfo {volume} {100}},\ \bibinfo
  {pages} {013617} (\bibinfo {year} {2019})}\BibitemShut {NoStop}%
\bibitem [{\citenamefont {Bai}\ \emph {et~al.}(2020)\citenamefont {Bai},
  \citenamefont {Gaur}, \citenamefont {Sable}, \citenamefont {Bandyopadhyay},
  \citenamefont {Suthar},\ and\ \citenamefont {Angom}}]{bai_20}%
  \BibitemOpen
  \bibfield  {author} {\bibinfo {author} {\bibfnamefont {R.}~\bibnamefont
  {Bai}}, \bibinfo {author} {\bibfnamefont {D.}~\bibnamefont {Gaur}}, \bibinfo
  {author} {\bibfnamefont {H.}~\bibnamefont {Sable}}, \bibinfo {author}
  {\bibfnamefont {S.}~\bibnamefont {Bandyopadhyay}}, \bibinfo {author}
  {\bibfnamefont {K.}~\bibnamefont {Suthar}},\ and\ \bibinfo {author}
  {\bibfnamefont {D.}~\bibnamefont {Angom}},\ }\href
  {https://doi.org/10.1103/PhysRevA.102.043309} {\bibfield  {journal} {\bibinfo
   {journal} {Phys. Rev. A}\ }\textbf {\bibinfo {volume} {102}},\ \bibinfo
  {pages} {043309} (\bibinfo {year} {2020})}\BibitemShut {NoStop}%
\bibitem [{\citenamefont {Ji}\ \emph {et~al.}(2014)\citenamefont {Ji},
  \citenamefont {Zhang}, \citenamefont {Zhang}, \citenamefont {Du},
  \citenamefont {Zheng}, \citenamefont {Deng}, \citenamefont {Zhai},
  \citenamefont {Chen},\ and\ \citenamefont {Pan}}]{ji_14}%
  \BibitemOpen
  \bibfield  {author} {\bibinfo {author} {\bibfnamefont {S.-C.}\ \bibnamefont
  {Ji}}, \bibinfo {author} {\bibfnamefont {J.-Y.}\ \bibnamefont {Zhang}},
  \bibinfo {author} {\bibfnamefont {L.}~\bibnamefont {Zhang}}, \bibinfo
  {author} {\bibfnamefont {Z.-D.}\ \bibnamefont {Du}}, \bibinfo {author}
  {\bibfnamefont {W.}~\bibnamefont {Zheng}}, \bibinfo {author} {\bibfnamefont
  {Y.-J.}\ \bibnamefont {Deng}}, \bibinfo {author} {\bibfnamefont
  {H.}~\bibnamefont {Zhai}}, \bibinfo {author} {\bibfnamefont {S.}~\bibnamefont
  {Chen}},\ and\ \bibinfo {author} {\bibfnamefont {J.-W.}\ \bibnamefont
  {Pan}},\ }\href {https://doi.org/10.1038/nphys2905} {\bibfield  {journal}
  {\bibinfo  {journal} {Nat. Phys.}\ }\textbf {\bibinfo {volume} {10}},\
  \bibinfo {pages} {314} (\bibinfo {year} {2014})}\BibitemShut {NoStop}%
\bibitem [{\citenamefont {Burdick}\ \emph {et~al.}(2016)\citenamefont
  {Burdick}, \citenamefont {Tang},\ and\ \citenamefont {Lev}}]{burdick_16}%
  \BibitemOpen
  \bibfield  {author} {\bibinfo {author} {\bibfnamefont {N.~Q.}\ \bibnamefont
  {Burdick}}, \bibinfo {author} {\bibfnamefont {Y.}~\bibnamefont {Tang}},\ and\
  \bibinfo {author} {\bibfnamefont {B.~L.}\ \bibnamefont {Lev}},\ }\href
  {https://doi.org/10.1103/PhysRevX.6.031022} {\bibfield  {journal} {\bibinfo
  {journal} {Phys. Rev. X}\ }\textbf {\bibinfo {volume} {6}},\ \bibinfo {pages}
  {031022} (\bibinfo {year} {2016})}\BibitemShut {NoStop}%
\bibitem [{\citenamefont {Norcia}\ and\ \citenamefont
  {Ferlaino}(2021)}]{norcia_21a}%
  \BibitemOpen
  \bibfield  {author} {\bibinfo {author} {\bibfnamefont {M.~A.}\ \bibnamefont
  {Norcia}}\ and\ \bibinfo {author} {\bibfnamefont {F.}~\bibnamefont
  {Ferlaino}},\ }\href {https://doi.org/10.1038/s41567-021-01398-7} {\bibfield
  {journal} {\bibinfo  {journal} {Nat. Phys.}\ }\textbf {\bibinfo {volume}
  {17}},\ \bibinfo {pages} {1349} (\bibinfo {year} {2021})}\BibitemShut
  {NoStop}%
\bibitem [{\citenamefont {Gong}\ \emph {et~al.}(2015)\citenamefont {Gong},
  \citenamefont {Qian}, \citenamefont {Yan}, \citenamefont {Scarola},\ and\
  \citenamefont {Zhang}}]{gong_15}%
  \BibitemOpen
  \bibfield  {author} {\bibinfo {author} {\bibfnamefont {M.}~\bibnamefont
  {Gong}}, \bibinfo {author} {\bibfnamefont {Y.}~\bibnamefont {Qian}}, \bibinfo
  {author} {\bibfnamefont {M.}~\bibnamefont {Yan}}, \bibinfo {author}
  {\bibfnamefont {V.~W.}\ \bibnamefont {Scarola}},\ and\ \bibinfo {author}
  {\bibfnamefont {C.}~\bibnamefont {Zhang}},\ }\href
  {https://doi.org/10.1038/srep10050} {\bibfield  {journal} {\bibinfo
  {journal} {Sci. Rep.}\ }\textbf {\bibinfo {volume} {5}},\ \bibinfo {pages}
  {10050} (\bibinfo {year} {2015})}\BibitemShut {NoStop}%
\bibitem [{\citenamefont {Cai}\ \emph {et~al.}(2012)\citenamefont {Cai},
  \citenamefont {Zhou},\ and\ \citenamefont {Wu}}]{cai_12}%
  \BibitemOpen
  \bibfield  {author} {\bibinfo {author} {\bibfnamefont {Z.}~\bibnamefont
  {Cai}}, \bibinfo {author} {\bibfnamefont {X.}~\bibnamefont {Zhou}},\ and\
  \bibinfo {author} {\bibfnamefont {C.}~\bibnamefont {Wu}},\ }\href
  {https://doi.org/10.1103/PhysRevA.85.061605} {\bibfield  {journal} {\bibinfo
  {journal} {Phys. Rev. A}\ }\textbf {\bibinfo {volume} {85}},\ \bibinfo
  {pages} {061605} (\bibinfo {year} {2012})}\BibitemShut {NoStop}%
\bibitem [{\citenamefont {Zhang}\ \emph {et~al.}(2019)\citenamefont {Zhang},
  \citenamefont {Ke},\ and\ \citenamefont {Lee}}]{zhang_19}%
  \BibitemOpen
  \bibfield  {author} {\bibinfo {author} {\bibfnamefont {L.}~\bibnamefont
  {Zhang}}, \bibinfo {author} {\bibfnamefont {Y.}~\bibnamefont {Ke}},\ and\
  \bibinfo {author} {\bibfnamefont {C.}~\bibnamefont {Lee}},\ }\href
  {https://doi.org/10.1103/PhysRevB.100.224420} {\bibfield  {journal} {\bibinfo
   {journal} {Phys. Rev. B}\ }\textbf {\bibinfo {volume} {100}},\ \bibinfo
  {pages} {224420} (\bibinfo {year} {2019})}\BibitemShut {NoStop}%
\bibitem [{\citenamefont {Cole}\ \emph {et~al.}(2012)\citenamefont {Cole},
  \citenamefont {Zhang}, \citenamefont {Paramekanti},\ and\ \citenamefont
  {Trivedi}}]{cole_12}%
  \BibitemOpen
  \bibfield  {author} {\bibinfo {author} {\bibfnamefont {W.~S.}\ \bibnamefont
  {Cole}}, \bibinfo {author} {\bibfnamefont {S.}~\bibnamefont {Zhang}},
  \bibinfo {author} {\bibfnamefont {A.}~\bibnamefont {Paramekanti}},\ and\
  \bibinfo {author} {\bibfnamefont {N.}~\bibnamefont {Trivedi}},\ }\href
  {https://doi.org/10.1103/PhysRevLett.109.085302} {\bibfield  {journal}
  {\bibinfo  {journal} {Phys. Rev. Lett.}\ }\textbf {\bibinfo {volume} {109}},\
  \bibinfo {pages} {085302} (\bibinfo {year} {2012})}\BibitemShut {NoStop}%
\bibitem [{\citenamefont {Sousa-J\'unior}\ and\ \citenamefont
  {Mondaini}(2025)}]{sousa_25}%
  \BibitemOpen
  \bibfield  {author} {\bibinfo {author} {\bibfnamefont {S.~a. d.~A.}\
  \bibnamefont {Sousa-J\'unior}}\ and\ \bibinfo {author} {\bibfnamefont
  {R.}~\bibnamefont {Mondaini}},\ }\href
  {https://doi.org/10.1103/PhysRevB.111.075166} {\bibfield  {journal} {\bibinfo
   {journal} {Phys. Rev. B}\ }\textbf {\bibinfo {volume} {111}},\ \bibinfo
  {pages} {075166} (\bibinfo {year} {2025})}\BibitemShut {NoStop}%
\bibitem [{\citenamefont {Bersano}\ \emph {et~al.}(2019)\citenamefont
  {Bersano}, \citenamefont {Hou}, \citenamefont {Mossman}, \citenamefont
  {Gokhroo}, \citenamefont {Luo}, \citenamefont {Sun}, \citenamefont {Zhang},\
  and\ \citenamefont {Engels}}]{bersano_19}%
  \BibitemOpen
  \bibfield  {author} {\bibinfo {author} {\bibfnamefont {T.~M.}\ \bibnamefont
  {Bersano}}, \bibinfo {author} {\bibfnamefont {J.}~\bibnamefont {Hou}},
  \bibinfo {author} {\bibfnamefont {S.}~\bibnamefont {Mossman}}, \bibinfo
  {author} {\bibfnamefont {V.}~\bibnamefont {Gokhroo}}, \bibinfo {author}
  {\bibfnamefont {X.-W.}\ \bibnamefont {Luo}}, \bibinfo {author} {\bibfnamefont
  {K.}~\bibnamefont {Sun}}, \bibinfo {author} {\bibfnamefont {C.}~\bibnamefont
  {Zhang}},\ and\ \bibinfo {author} {\bibfnamefont {P.}~\bibnamefont
  {Engels}},\ }\href {https://doi.org/10.1103/PhysRevA.99.051602} {\bibfield
  {journal} {\bibinfo  {journal} {Phys. Rev. A}\ }\textbf {\bibinfo {volume}
  {99}},\ \bibinfo {pages} {051602} (\bibinfo {year} {2019})}\BibitemShut
  {NoStop}%
\bibitem [{\citenamefont {Gra\ss{}}\ \emph {et~al.}(2011)\citenamefont
  {Gra\ss{}}, \citenamefont {Saha}, \citenamefont {Sengupta},\ and\
  \citenamefont {Lewenstein}}]{grass_11}%
  \BibitemOpen
  \bibfield  {author} {\bibinfo {author} {\bibfnamefont {T.}~\bibnamefont
  {Gra\ss{}}}, \bibinfo {author} {\bibfnamefont {K.}~\bibnamefont {Saha}},
  \bibinfo {author} {\bibfnamefont {K.}~\bibnamefont {Sengupta}},\ and\
  \bibinfo {author} {\bibfnamefont {M.}~\bibnamefont {Lewenstein}},\ }\href
  {https://doi.org/10.1103/PhysRevA.84.053632} {\bibfield  {journal} {\bibinfo
  {journal} {Phys. Rev. A}\ }\textbf {\bibinfo {volume} {84}},\ \bibinfo
  {pages} {053632} (\bibinfo {year} {2011})}\BibitemShut {NoStop}%
\bibitem [{\citenamefont {Mandal}\ \emph {et~al.}(2012)\citenamefont {Mandal},
  \citenamefont {Saha},\ and\ \citenamefont {Sengupta}}]{mandal_12}%
  \BibitemOpen
  \bibfield  {author} {\bibinfo {author} {\bibfnamefont {S.}~\bibnamefont
  {Mandal}}, \bibinfo {author} {\bibfnamefont {K.}~\bibnamefont {Saha}},\ and\
  \bibinfo {author} {\bibfnamefont {K.}~\bibnamefont {Sengupta}},\ }\href
  {https://doi.org/10.1103/PhysRevB.86.155101} {\bibfield  {journal} {\bibinfo
  {journal} {Phys. Rev. B}\ }\textbf {\bibinfo {volume} {86}},\ \bibinfo
  {pages} {155101} (\bibinfo {year} {2012})}\BibitemShut {NoStop}%
\bibitem [{\citenamefont {Hickey}\ and\ \citenamefont
  {Paramekanti}(2014)}]{hickey_14}%
  \BibitemOpen
  \bibfield  {author} {\bibinfo {author} {\bibfnamefont {C.}~\bibnamefont
  {Hickey}}\ and\ \bibinfo {author} {\bibfnamefont {A.}~\bibnamefont
  {Paramekanti}},\ }\href {https://doi.org/10.1103/PhysRevLett.113.265302}
  {\bibfield  {journal} {\bibinfo  {journal} {Phys. Rev. Lett.}\ }\textbf
  {\bibinfo {volume} {113}},\ \bibinfo {pages} {265302} (\bibinfo {year}
  {2014})}\BibitemShut {NoStop}%
\bibitem [{\citenamefont {Dutta}\ \emph {et~al.}(2019)\citenamefont {Dutta},
  \citenamefont {Joshi}, \citenamefont {Sengupta},\ and\ \citenamefont
  {Majumdar}}]{dutta_19}%
  \BibitemOpen
  \bibfield  {author} {\bibinfo {author} {\bibfnamefont {A.}~\bibnamefont
  {Dutta}}, \bibinfo {author} {\bibfnamefont {A.}~\bibnamefont {Joshi}},
  \bibinfo {author} {\bibfnamefont {K.}~\bibnamefont {Sengupta}},\ and\
  \bibinfo {author} {\bibfnamefont {P.}~\bibnamefont {Majumdar}},\ }\href
  {https://doi.org/10.1103/PhysRevB.99.195126} {\bibfield  {journal} {\bibinfo
  {journal} {Phys. Rev. B}\ }\textbf {\bibinfo {volume} {99}},\ \bibinfo
  {pages} {195126} (\bibinfo {year} {2019})}\BibitemShut {NoStop}%
\bibitem [{\citenamefont {Suthar}\ \emph {et~al.}(2021)\citenamefont {Suthar},
  \citenamefont {Kaur}, \citenamefont {Gautam},\ and\ \citenamefont
  {Angom}}]{suthar_21}%
  \BibitemOpen
  \bibfield  {author} {\bibinfo {author} {\bibfnamefont {K.}~\bibnamefont
  {Suthar}}, \bibinfo {author} {\bibfnamefont {P.}~\bibnamefont {Kaur}},
  \bibinfo {author} {\bibfnamefont {S.}~\bibnamefont {Gautam}},\ and\ \bibinfo
  {author} {\bibfnamefont {D.}~\bibnamefont {Angom}},\ }\href
  {https://doi.org/10.1103/PhysRevA.104.043320} {\bibfield  {journal} {\bibinfo
   {journal} {Phys. Rev. A}\ }\textbf {\bibinfo {volume} {104}},\ \bibinfo
  {pages} {043320} (\bibinfo {year} {2021})}\BibitemShut {NoStop}%
\bibitem [{\citenamefont {Pu}\ \emph {et~al.}(2024)\citenamefont {Pu},
  \citenamefont {Wang}, \citenamefont {Song},\ and\ \citenamefont
  {Bai}}]{pu_24}%
  \BibitemOpen
  \bibfield  {author} {\bibinfo {author} {\bibfnamefont {D.-D.}\ \bibnamefont
  {Pu}}, \bibinfo {author} {\bibfnamefont {J.-G.}\ \bibnamefont {Wang}},
  \bibinfo {author} {\bibfnamefont {Y.-F.}\ \bibnamefont {Song}},\ and\
  \bibinfo {author} {\bibfnamefont {X.-D.}\ \bibnamefont {Bai}},\ }\href
  {https://doi.org/10.1088/1367-2630/ad388e} {\bibfield  {journal} {\bibinfo
  {journal} {New J. Phys.}\ }\textbf {\bibinfo {volume} {26}},\ \bibinfo
  {pages} {043003} (\bibinfo {year} {2024})}\BibitemShut {NoStop}%
\bibitem [{\citenamefont {Giovanazzi}\ \emph {et~al.}(2002)\citenamefont
  {Giovanazzi}, \citenamefont {G\"orlitz},\ and\ \citenamefont
  {Pfau}}]{giovanazzi_02}%
  \BibitemOpen
  \bibfield  {author} {\bibinfo {author} {\bibfnamefont {S.}~\bibnamefont
  {Giovanazzi}}, \bibinfo {author} {\bibfnamefont {A.}~\bibnamefont
  {G\"orlitz}},\ and\ \bibinfo {author} {\bibfnamefont {T.}~\bibnamefont
  {Pfau}},\ }\href {https://doi.org/10.1103/PhysRevLett.89.130401} {\bibfield
  {journal} {\bibinfo  {journal} {Phys. Rev. Lett.}\ }\textbf {\bibinfo
  {volume} {89}},\ \bibinfo {pages} {130401} (\bibinfo {year}
  {2002})}\BibitemShut {NoStop}%
\bibitem [{\citenamefont {Tang}\ \emph {et~al.}(2018)\citenamefont {Tang},
  \citenamefont {Kao}, \citenamefont {Li},\ and\ \citenamefont
  {Lev}}]{tang_18}%
  \BibitemOpen
  \bibfield  {author} {\bibinfo {author} {\bibfnamefont {Y.}~\bibnamefont
  {Tang}}, \bibinfo {author} {\bibfnamefont {W.}~\bibnamefont {Kao}}, \bibinfo
  {author} {\bibfnamefont {K.-Y.}\ \bibnamefont {Li}},\ and\ \bibinfo {author}
  {\bibfnamefont {B.~L.}\ \bibnamefont {Lev}},\ }\href
  {https://doi.org/10.1103/PhysRevLett.120.230401} {\bibfield  {journal}
  {\bibinfo  {journal} {Phys. Rev. Lett.}\ }\textbf {\bibinfo {volume} {120}},\
  \bibinfo {pages} {230401} (\bibinfo {year} {2018})}\BibitemShut {NoStop}%
\bibitem [{\citenamefont {Gutzwiller}(1963)}]{gutzwiller_63}%
  \BibitemOpen
  \bibfield  {author} {\bibinfo {author} {\bibfnamefont {M.~C.}\ \bibnamefont
  {Gutzwiller}},\ }\href {https://doi.org/10.1103/PhysRevLett.10.159}
  {\bibfield  {journal} {\bibinfo  {journal} {Phys. Rev. Lett.}\ }\textbf
  {\bibinfo {volume} {10}},\ \bibinfo {pages} {159} (\bibinfo {year}
  {1963})}\BibitemShut {NoStop}%
\bibitem [{\citenamefont {Rokhsar}\ and\ \citenamefont
  {Kotliar}(1991)}]{rokshar_91}%
  \BibitemOpen
  \bibfield  {author} {\bibinfo {author} {\bibfnamefont {D.~S.}\ \bibnamefont
  {Rokhsar}}\ and\ \bibinfo {author} {\bibfnamefont {B.~G.}\ \bibnamefont
  {Kotliar}},\ }\href {https://doi.org/10.1103/PhysRevB.44.10328} {\bibfield
  {journal} {\bibinfo  {journal} {Phys. Rev. B}\ }\textbf {\bibinfo {volume}
  {44}},\ \bibinfo {pages} {10328} (\bibinfo {year} {1991})}\BibitemShut
  {NoStop}%
\bibitem [{\citenamefont {Sheshadri}\ \emph {et~al.}(1993)\citenamefont
  {Sheshadri}, \citenamefont {Krishnamurthy}, \citenamefont {Pandit},\ and\
  \citenamefont {Ramakrishnan}}]{sheshadri_93}%
  \BibitemOpen
  \bibfield  {author} {\bibinfo {author} {\bibfnamefont {K.}~\bibnamefont
  {Sheshadri}}, \bibinfo {author} {\bibfnamefont {H.~R.}\ \bibnamefont
  {Krishnamurthy}}, \bibinfo {author} {\bibfnamefont {R.}~\bibnamefont
  {Pandit}},\ and\ \bibinfo {author} {\bibfnamefont {T.~V.}\ \bibnamefont
  {Ramakrishnan}},\ }\href {https://doi.org/10.1209/0295-5075/22/4/004}
  {\bibfield  {journal} {\bibinfo  {journal} {EPL}\ }\textbf {\bibinfo {volume}
  {22}},\ \bibinfo {pages} {257} (\bibinfo {year} {1993})}\BibitemShut
  {NoStop}%
\bibitem [{\citenamefont {Bai}\ \emph {et~al.}(2018)\citenamefont {Bai},
  \citenamefont {Bandyopadhyay}, \citenamefont {Pal}, \citenamefont {Suthar},\
  and\ \citenamefont {Angom}}]{bai_18}%
  \BibitemOpen
  \bibfield  {author} {\bibinfo {author} {\bibfnamefont {R.}~\bibnamefont
  {Bai}}, \bibinfo {author} {\bibfnamefont {S.}~\bibnamefont {Bandyopadhyay}},
  \bibinfo {author} {\bibfnamefont {S.}~\bibnamefont {Pal}}, \bibinfo {author}
  {\bibfnamefont {K.}~\bibnamefont {Suthar}},\ and\ \bibinfo {author}
  {\bibfnamefont {D.}~\bibnamefont {Angom}},\ }\href
  {https://doi.org/10.1103/PhysRevA.98.023606} {\bibfield  {journal} {\bibinfo
  {journal} {Phys. Rev. A}\ }\textbf {\bibinfo {volume} {98}},\ \bibinfo
  {pages} {023606} (\bibinfo {year} {2018})}\BibitemShut {NoStop}%
\bibitem [{\citenamefont {Suthar}\ \emph {et~al.}(2020)\citenamefont {Suthar},
  \citenamefont {Sable}, \citenamefont {Bai}, \citenamefont {Bandyopadhyay},
  \citenamefont {Pal},\ and\ \citenamefont {Angom}}]{suthar_20}%
  \BibitemOpen
  \bibfield  {author} {\bibinfo {author} {\bibfnamefont {K.}~\bibnamefont
  {Suthar}}, \bibinfo {author} {\bibfnamefont {H.}~\bibnamefont {Sable}},
  \bibinfo {author} {\bibfnamefont {R.}~\bibnamefont {Bai}}, \bibinfo {author}
  {\bibfnamefont {S.}~\bibnamefont {Bandyopadhyay}}, \bibinfo {author}
  {\bibfnamefont {S.}~\bibnamefont {Pal}},\ and\ \bibinfo {author}
  {\bibfnamefont {D.}~\bibnamefont {Angom}},\ }\href
  {https://doi.org/10.1103/PhysRevA.102.013320} {\bibfield  {journal} {\bibinfo
   {journal} {Phys. Rev. A}\ }\textbf {\bibinfo {volume} {102}},\ \bibinfo
  {pages} {013320} (\bibinfo {year} {2020})}\BibitemShut {NoStop}%
\bibitem [{\citenamefont {Suthar}\ and\ \citenamefont {Ng}(2022)}]{suthar_22}%
  \BibitemOpen
  \bibfield  {author} {\bibinfo {author} {\bibfnamefont {K.}~\bibnamefont
  {Suthar}}\ and\ \bibinfo {author} {\bibfnamefont {K.-K.}\ \bibnamefont
  {Ng}},\ }\href {https://doi.org/10.1103/PhysRevA.106.063313} {\bibfield
  {journal} {\bibinfo  {journal} {Phys. Rev. A}\ }\textbf {\bibinfo {volume}
  {106}},\ \bibinfo {pages} {063313} (\bibinfo {year} {2022})}\BibitemShut
  {NoStop}%
\bibitem [{\citenamefont {van Otterlo}\ and\ \citenamefont
  {Wagenblast}(1994)}]{otterlo_94}%
  \BibitemOpen
  \bibfield  {author} {\bibinfo {author} {\bibfnamefont {A.}~\bibnamefont {van
  Otterlo}}\ and\ \bibinfo {author} {\bibfnamefont {K.-H.}\ \bibnamefont
  {Wagenblast}},\ }\href {https://doi.org/10.1103/PhysRevLett.72.3598}
  {\bibfield  {journal} {\bibinfo  {journal} {Phys. Rev. Lett.}\ }\textbf
  {\bibinfo {volume} {72}},\ \bibinfo {pages} {3598} (\bibinfo {year}
  {1994})}\BibitemShut {NoStop}%
\bibitem [{\citenamefont {van Oosten}\ \emph {et~al.}(2001)\citenamefont {van
  Oosten}, \citenamefont {van~der Straten},\ and\ \citenamefont
  {Stoof}}]{oosten_01}%
  \BibitemOpen
  \bibfield  {author} {\bibinfo {author} {\bibfnamefont {D.}~\bibnamefont {van
  Oosten}}, \bibinfo {author} {\bibfnamefont {P.}~\bibnamefont {van~der
  Straten}},\ and\ \bibinfo {author} {\bibfnamefont {H.~T.~C.}\ \bibnamefont
  {Stoof}},\ }\href {https://doi.org/10.1103/PhysRevA.63.053601} {\bibfield
  {journal} {\bibinfo  {journal} {Phys. Rev. A}\ }\textbf {\bibinfo {volume}
  {63}},\ \bibinfo {pages} {053601} (\bibinfo {year} {2001})}\BibitemShut
  {NoStop}%
\bibitem [{\citenamefont {Iskin}\ and\ \citenamefont
  {Freericks}(2009)}]{iskin_09}%
  \BibitemOpen
  \bibfield  {author} {\bibinfo {author} {\bibfnamefont {M.}~\bibnamefont
  {Iskin}}\ and\ \bibinfo {author} {\bibfnamefont {J.~K.}\ \bibnamefont
  {Freericks}},\ }\href {https://doi.org/10.1103/PhysRevA.79.053634} {\bibfield
   {journal} {\bibinfo  {journal} {Phys. Rev. A}\ }\textbf {\bibinfo {volume}
  {79}},\ \bibinfo {pages} {053634} (\bibinfo {year} {2009})}\BibitemShut
  {NoStop}%
\bibitem [{\citenamefont {Yan}\ \emph {et~al.}(2017)\citenamefont {Yan},
  \citenamefont {Qian}, \citenamefont {Hui}, \citenamefont {Gong},
  \citenamefont {Zhang},\ and\ \citenamefont {Scarola}}]{yan_17}%
  \BibitemOpen
  \bibfield  {author} {\bibinfo {author} {\bibfnamefont {M.}~\bibnamefont
  {Yan}}, \bibinfo {author} {\bibfnamefont {Y.}~\bibnamefont {Qian}}, \bibinfo
  {author} {\bibfnamefont {H.-Y.}\ \bibnamefont {Hui}}, \bibinfo {author}
  {\bibfnamefont {M.}~\bibnamefont {Gong}}, \bibinfo {author} {\bibfnamefont
  {C.}~\bibnamefont {Zhang}},\ and\ \bibinfo {author} {\bibfnamefont {V.~W.}\
  \bibnamefont {Scarola}},\ }\href {https://doi.org/10.1103/PhysRevA.96.053619}
  {\bibfield  {journal} {\bibinfo  {journal} {Phys. Rev. A}\ }\textbf {\bibinfo
  {volume} {96}},\ \bibinfo {pages} {053619} (\bibinfo {year}
  {2017})}\BibitemShut {NoStop}%
\end{thebibliography}%
\bibliographystyle{apsrev4-2}
\end{document}